\documentclass[12pt,a4paper]{article}
\usepackage{epsfig}
\usepackage[a4paper]{geometry}
\usepackage{float}
\usepackage{mathrsfs}
\usepackage[stable]{footmisc}
\usepackage[utf8]{inputenc}
\usepackage{a4wide}
\usepackage{amsmath,amssymb,amstext,amsthm,amssymb}
\usepackage{amsfonts}
\usepackage{framed}
\usepackage{graphicx}
\usepackage[usenames, dvipsnames, table]{xcolor}
\usepackage{euscript}
\usepackage{color}
\usepackage{bbold}
\usepackage[normalem]{ulem}
\usepackage{bbm}
\usepackage{soul}
\usepackage{rotating} 
\usepackage{slashed} 
\usepackage{cancel} 
\usepackage{braket} 
\usepackage{amstext}
\usepackage{array}
\usepackage{setspace}
\usepackage{hyperref}
\usepackage{overpic}
\usepackage{bbm}
\usepackage{caption}
\usepackage{subcaption}
\usepackage{cite}
\usepackage{ulem}
\usepackage{lipsum}
\usepackage{longtable}
\usepackage{tikz}
\usepackage{enumitem}

\newcommand{\df}{\mathrel{:=}}

\usepackage{diagbox}
\newcommand\mathdiagbox[3][]{\hbox{\tabcolsep=\arraycolsep\diagbox[#1]{$#2$}{$#3$}}}

\makeatletter
\def\cstar#1{\expandafter\@cstar\csname c@#1\endcsname}
\def\@cstar#1{\ifcase#1\or $^*$\or $^*$\or $^*$\fi}
\AddEnumerateCounter{\cstar}{\@cstar}{$^*$}
\makeatother

\newcommand{\DeclareRuneSeparators}[1]{} 

\input{arm.fd}

\newcommand{\del}[0]{\partial}

\let\baraccent=\=
\renewcommand{\=}[1]{\stackrel{#1}{=}}

\DeclareSymbolFontAlphabet{\mathbb}{AMSb}

\DeclareMathOperator{\Vis}{Vis}

\makeatletter
\@addtoreset{equation}{section}
\makeatother

\begin{document}
	

	
	\pagestyle{empty}
	
	{\hfill ACFI-T22-10}
	
	\begin{center}
  {\LARGE \bf{Moduli Space Reconstruction\\
   and
   Weak Gravity} \\[4mm] }
   {\large \bf or \\[4mm] Unto the Ends of the Moduli Space} \\[6mm]
   	With Illustrations by the Authors \\[15mm]
	\end{center}
	
	
	\begin{center}
		\scalebox{0.95}[0.95]{{\fontsize{14}{30}\selectfont Naomi Gendler,$^{a,b}$ Ben Heidenreich,$^{c}$ Liam McAllister,$^{a}$}} \vspace{0.35cm}
		\scalebox{0.95}[0.95]{{\fontsize{14}{30}\selectfont Jakob Moritz,$^{a,d}$ and Tom Rudelius$^{e,f}$}}
	\end{center}
	
	\begin{center}
		\vspace{0.25 cm}
		\textsl{$^{a}$Department of Physics, Cornell University, Ithaca, NY 14853, USA}\\
		\textsl{$^{b}$Department of Physics, Harvard University, Cambridge, MA 02138, USA}\\
		\textsl{$^{c}$Department of Physics, University of Massachusetts, Amherst, MA 01003, USA}\\
		\textsl{$^{d}$Department of Theoretical Physics, CERN, 1211 Meyrin, Switzerland}\\
		\textsl{$^{e}$Department of Physics, University of California, Berkeley, CA 94720, USA}\\
		\textsl{$^{f}$Department of Mathematical Sciences, Durham University, Durham DH1 3LE, UK}\\
		
		\vspace{1cm}
		\normalsize{\bf Abstract} \\[8mm]

	\end{center}

We present a method to construct the extended K\"ahler cone of any Calabi-Yau threefold by using Gopakumar-Vafa invariants to identify all geometric phases that are related by flops or Weyl reflections.
In this way we obtain the K\"ahler moduli spaces of all favorable Calabi-Yau threefold hypersurfaces with $h^{1,1} \le 4$, including
toric and non-toric phases.
In this setting we perform an explicit test of the Weak Gravity Conjecture by using the Gopakumar-Vafa invariants to count BPS states.  All of our examples satisfy the tower/sublattice WGC, and in fact they even satisfy the stronger lattice WGC.

	\begin{center}
		\begin{minipage}[h]{2cm}

		\end{minipage}
	\end{center}
	\newpage
	\setcounter{page}{1}
	\pagestyle{plain}
	\renewcommand{\thefootnote}{\arabic{footnote}}
	\setcounter{footnote}{0}
	%
	%
	\tableofcontents
	\newpage

\section{Introduction}
\label{sec:intro}

The
physics resulting from a compactification of string theory
often varies across a moduli space of geometries.
Characterizing the moduli space is then a prerequisite for understanding the imprint of quantum gravity on low-energy physics.
In general the moduli space is the union of a number of regions, for example corresponding to distinct geometric phases, and a primary task is to map out the transitions among these components.

In this work we will consider the K\"ahler moduli space of birationally-equivalent Calabi-Yau threefolds.
This moduli space is a cone, the extended K\"ahler cone $\mathcal{K}$, whose dimension is given by the Hodge number $h^{1,1}$ of the threefold.
The extended K\"ahler cone is the union of a collection of smaller cones, each corresponding to the K\"ahler cone of a distinct geometric phase, and linked to its neighbors by birational transformations: see Figure \ref{fig:extended_kahler}.  The number of phases grows rapidly as $h^{1,1}$ increases, so much so that constructing $\mathcal{K}$ becomes challenging for $h^{1,1} > 2$.

The first result of this paper is a general algorithm for constructing the extended K\"ahler cone.
We probe the Calabi-Yau via compactification of M-theory.
Starting from one phase, one needs to know which other phases to adjoin via flop transitions.  We show how to use the Gopakumar-Vafa (GV) invariants of curves to identify the set of curves that can be
flopped. In the process we also identify curves associated to $\mathfrak{su}(2)$ enhancements, and we show how to account for the gauge redundancy resulting from their
Weyl reflections. We then use the data of flops and Weyl reflections to map out the entire K\"ahler moduli space.\footnote{As explained in \cite{Alim:2021vhs}, the data of the extended K\"ahler cone also provides a means of studying the complete effective cones in these geometries, but we defer this application to future work.}

In the case of Calabi-Yau threefold hypersurfaces in toric varieties, one can compute the GV invariants by means of mirror symmetry \cite{Hosono:1993qy,Hosono:1994ax}, for example using {\tt{CYTools}} \cite{Demirtas:2022hqf,computational-mirror-symmetry}.  In this setting we
construct the K\"ahler moduli spaces of all favorable geometries with $h^{1,1} \le 4$, and of a selection of geometries with $h^{1,1}=5$.

The second result of this paper is an application of the first: we test the Weak Gravity Conjecture (WGC)~\cite{Arkani-Hamed:2006emk} in our ensemble of compactifications
on Calabi-Yau threefold hypersurfaces by reconstructing the K\"ahler moduli space in each case.
The WGC predicts that certain superextremal states should exist.
In particular, in the region in charge space where BPS black holes\footnote{These black holes need not be spherically symmetric, and could in fact be multi-centered.} exist, denoted hereafter by $\mathscr{C}_{\mathrm{BH}}$, the WGC predicts the existence of BPS particles.\footnote{Specifically, the lattice WGC predicts the existence of a BPS particle for each site in the intersection of the charge lattice with $\mathscr{C}_{\mathrm{BH}}$;
the sublattice WGC predicts the existence of a BPS particle on a full-dimensional sublattice in $\mathscr{C}_{\mathrm{BH}}$; and the tower WGC predicts an infinite tower of BPS particles along each rational ray in $\mathscr{C}_{\mathrm{BH}}$.}
The GV invariants are BPS indices, and so a sufficient --- but not necessary --- condition for the existence of a  BPS particle of a given electric charge is that the corresponding GV invariant is nonzero. Building on \cite{Alim:2021vhs}, we show how to compute a region contained in $\mathscr{C}_{\mathrm{BH}}$ by carrying out all flops, as well as all Weyl reflections across which all BPS states remain stable.
Upon computing this region and the GV invariants of charges therein, we find that the tower and sublattice WGC \cite{Heidenreich:2016aqi, Andriolo:2018lvp}, and even the stronger lattice WGC \cite{Heidenreich:2015nta}, are obeyed in all our examples, out to the highest charges we examined. This is somewhat surprising because there are known counterexamples to the lattice WGC \cite{Heidenreich:2016aqi}, which arise in certain toroidal compactifications of type II/heterotic string theory. It is significant, perhaps, that no such counterexamples arise in the class of Calabi-Yau compactifications considered here.

The structure of this paper is as follows. In \S\ref{sec:review} we review some salient aspects of Calabi-Yau geometry.
In \S\ref{sec:modulispace} we explain the transitions that can occur when passing through a wall in the K\"ahler moduli space.
We show how to use the data of GV invariants to identify curves that can be flopped, and thus connect birationally-equivalent Calabi-Yau threefolds.
We present an algorithm to assemble the extended K\"ahler cone by combining all possible flops, unto the ends of the moduli space.
In \S\ref{sec:wgc} we construct the extended K\"ahler cones of an ensemble of Calabi-Yau threefold hypersurfaces and use this knowledge
to carry out a large-scale test of the WGC.
We conclude in \S\ref{sec:conclusions}.

\section{Review of Calabi-Yau Compactifications} \label{sec:review}

In this section we will briefly review a few results from compactifications of M-theory and type IIA string theory on Calabi-Yau threefolds, focusing on the definitions and interrelations of a number of cones, and on their connections to Gopakumar-Vafa invariants.

\subsection{Effective theory}

We begin by compactifying M-theory on a Calabi-Yau threefold $X$. The result is a five-dimensional $\mathcal{N}=1$ effective supergravity theory. The massless particle spectrum is given by the gravity multiplet, $h^{2,1}(X)+1$ hypermultiplets, and $n:=h^{1,1}(X)-1$ vector multiplets. The $4h^{2,1}(X)+4$ scalar degrees of freedom in the hypermultiplets are furnished by the $2h^{2,1}(X)$ complex structure moduli of $X$, the $2h^{2,1}+3$ axions from dimensional reduction of the eleven-dimensional three-form, and the overall volume modulus of $X$, while the $n$ scalars parameterizing the vector multiplet moduli space come from the K\"ahler moduli of $X$ that leave the overall volume fixed. At generic points in K\"ahler moduli space the gauge group is $U(1)^{n+1}$, where $n$ of the abelian vector fields live in the $n$ vector multiplets and one lives in the gravity multiplet.

A general K\"ahler form $J$ can be expanded in a basis of divisor classes $\bigl\{[H_a]\bigr\}_{a=1}^{h^{1,1}(X)}$,
\begin{equation}\label{eq:Kahler_form}
	J=\sum_{a=1}^{h^{1,1}(X)}t^a[H_a]\, ,
\end{equation}
and one can think of the K\"ahler parameters $t^a$ as projective coordinates on vector multiplet moduli space. By
 a choice of normalization, integrating $J$ over effective curves in $X$ yields curve volumes in units of the eleven-dimensional Planck length $\ell_{11}$.\footnote{We define $\ell_{11}$ via the eleven-dimensional Einstein-Hilbert action $S_{EH}=\frac{2\pi}{\ell_{11}^9}\int \star R$.}

Neglecting the hypermultiplets, the bosonic action can be written as
\begin{align}
S =&\frac{2\pi}{\ell_5^3} \int d^5 x \sqrt{-g} \left(R - \frac{1}{2} \mathfrak{g}_{ij}(\phi) \partial \phi^i \cdot \partial \phi^j \right)\nonumber\\ &- \frac{1}{4\pi\ell_5} \int f_{ab}(\phi)F^a \wedge \star F^b
- \frac{1}{24\pi^2} \int \kappa_{abc} A^a \wedge F^b \wedge F^c\, ,
\label{eq:Mtheoryaction}
\end{align}
where the $\phi^i$, $i = 1, \ldots, n$,  are affine coordinates on moduli space, and the $A^a$ are the $n+1$ gauge potentials, with field strengths $F^a=dA^a$. Moreover, the $\kappa_{abc}:=\int_X [H_a]\wedge [H_b]\wedge [H_c]$ are the triple intersection numbers, and $\ell_5$ is the five-dimensional Planck length.

The scalar and gauge field kinetic terms are obtained by dimensional reduction of the eleven-dimensional effective action and can be written in terms of a homogeneous degree-three prepotential in the projective coordinates $t^a$,
\begin{equation}\label{eq:mtheoryprepotential}
\mathcal{F}^M[t]=\frac{1}{3!}\kappa_{abc} t^a t^b t^c\,.
\end{equation}
We may choose the $\phi^i$
as coordinates of the $n$-dimensional hypersurface $\mathcal{F}^M[t]=1$, thus gauge fixing the projective equivalence $t^a\sim \lambda t^a$. The kinetic terms are then
\begin{align}
f_{ab} &= \mathcal{F}^M_a \mathcal{F}^M_b - \mathcal{F}^M_{ab}\,, \quad
\mathfrak{g}_{ij} = f_{ab} \partial_i t^a \partial_j t^b\, ,
\end{align}
where $\mathcal{F}^{M}_a:=\del_a \mathcal{F}^M$ and $\mathcal{F}^M_{ab}:=\del_a\del_b \mathcal{F}^M$.

Importantly, as the overall volume modulus is in a hypermultiplet, the full prepotential $\mathcal{F}^M[t]$ can be computed in the large volume limit where it is classical, and thus receives neither perturbative nor non-perturbative corrections. This allows one to completely classify the possible behavior of the EFT at boundaries of the moduli space  \cite{Witten:1996qb}, as we review in \S\ref{sec:facets}.

We can also compactify type IIA string theory on $X$, which is equivalent to reducing the above five-dimensional theory on a circle. The result is a four-dimensional $\mathcal{N}=2$ effective supergravity theory, including $h^{2,1}(X)+1$ hypermultiplets directly inherited from the five-dimensional theory.  The $n$ five-dimensional vector multiplets descend to four-dimensional vector multiplets, and the reduction of the five-dimensional gravity multiplet to four dimensions provides one more vector multiplet, for $h^{1,1}$ in total.

The bosonic action, again neglecting hypermultiplets, is
\begin{align}
S = &\frac{2\pi}{\ell_4^2}\int d^4 x \sqrt{-g} \left(R -2K_{a\bar{b}}\del z^a \del \overline{z^b}\right)\nonumber\\
&+\frac{1}{4\pi} \int \left(\text{Im}(\mathcal{N}_{AB})F^A\wedge \star F^B +\text{Re}(\mathcal{N}_{AB})F^A\wedge  F^B\right)\, ,
\end{align}
where $A,B = 0, \ldots, h^{1,1}$ and $a,b = 1, \ldots, h^{1,1}$. The K\"ahler metric $K_{a\bar{b}}:=\del_a \overline{\del_b} K$ and the gauge-kinetic matrix $\mathcal{N}_{AB}$ can be written in terms of a holomorphic prepotential $\mathcal{F}^{IIA}$, which is homogeneous of degree two in projective coordinates $Z^A$:
\begin{align} \label{eq:defkahler}
K &= -\log \left(i \left[\overline{Z^A} \mathcal{F}^{\text{IIA}}_{A} - Z^A \overline{\mathcal{F}^{\text{IIA}}_A}\right]\right)\, , \\
\mathcal{N}_{AB} &= \overline{\mathcal{F}^{\text{IIA}}_{AB}} + 2i \frac{\text{Im}(\mathcal{F}^{\text{IIA}}_{AC})Z^C \text{Im}(\mathcal{F}^{\text{IIA}}_{BD})  Z^D}{\text{Im} (\mathcal{F}^{\text{IIA}}_{MN}) Z^M Z^N}\, .
\end{align}
The type IIA prepotential can be expanded around large volume as
\begin{align}
\frac{\mathcal{F}^{\text{IIA}}[Z]}{(Z^0)^2} = - \frac{1}{3!} \kappa_{abc} z^a z^b z^c &+ \frac{1}{2} a_{ab} z^a z^b + \frac{1}{24} c_a z^a + \frac{\zeta(3) \chi(X)}{2(2\pi i )^3} \nonumber\\&- \frac{1}{(2\pi i)^3}\sum_{[\mathcal{C}] \in \mathcal{M}_X} n_{[\mathcal{C}]}^0 \mathrm{Li}_3(q^{[\mathcal{C}]})
\label{eq:prepotentialIIA}
\end{align}
where $z^a:=Z^a/Z^0$ are $h^{1,1}$ complex affine coordinates on K\"ahler moduli space. The real parts $b^a:=\text{Re}(z^a)$ are the integrals of the ten-dimensional two-form $B_2$ over a basis of $H_2(X,\mathbb{Z})$, and the imaginary parts $\tilde{t}^a:=\text{Im}(z^a)$ measure string frame curve volumes.\footnote{The string frame curve volumes $\tilde{t}^a$ are related to the curve volumes $t^a$ in eleven-dimensional Planck units via $\tilde{t}^a=g_s^{\frac{2}{3}}t^a\equiv \text{Vol}(S^1)/\ell_{11} t^a$, with $g_s$ the type IIA string coupling.}

As in \eqref{eq:mtheoryprepotential}, $\kappa_{abc}$ are the triple intersection numbers of $X$.  The leading term in \eqref{eq:prepotentialIIA} is inherited classically from the five-dimensional prepotential, but in four dimensions there are also perturbative corrections in $\alpha'$ parameterized by
\begin{align}
c_a = \int_X c_2(X) \wedge [H_a], \ \ \ a_{ab} = \frac{1}{2}\,\begin{cases} \kappa_{aab} & a \geq b \\
\kappa_{abb} & a < b
\end{cases}\,, \ \ \text{and} \ \ \chi(X) = \int_X c_3(X),
\end{align}
with $c_2(X)$ and $c_3(X)$ the second and third Chern classes of $X$, respectively. Furthermore, the perturbative result is corrected by a series of worldsheet instanton corrections from strings wrapping effective curves in $X$, or equivalently BPS particles in the five-dimensional theory traveling around the compactification circle. The coefficients $n_{[\mathcal{C}]}^0$ are the BPS indices of the five-dimensional theory, namely genus zero GV invariants, and
\begin{equation}
	q^{[\mathcal{C}]}:=\exp\left(2\pi i \int_{\mathcal{C}}J^\mathbb{C}_{IIA}\right)\, ,
\end{equation}
where $J^{\mathbb{C}}_{IIA}:=B_2+i J_{IIA}$ and  $J_{IIA}=\sum_a \tilde{t}^a [H_a]$ is the K\"ahler form measuring string frame volumes. Decompactifying the four-dimensional theory back to five dimensions corresponds to taking the limit
\begin{equation}
	\tilde{t}^a\rightarrow \lambda^2 \tilde{t}^a\, ,\quad g_s\rightarrow \lambda^3 g_s\, ,\quad \lambda\rightarrow \infty\, ,
\end{equation}
and indeed one recovers the five-dimensional prepotential $\mathcal{F}^M[t^a]=-i \lim\limits_{\lambda\rightarrow \infty}\frac{\mathcal{F}^{\text{IIA}}[Z]}{g_s^2(Z^0)^2}$.

\subsection{Calabi-Yau geometry and BPS states} \label{sec:cygeometry}

Let us review a few well-known facts about Calabi-Yau threefolds and their K\"ahler moduli spaces. First, we recall the following theorem:

\vspace{.2cm}
\noindent
\textbf{Wall's theorem \cite{Wall}:} The diffeomorphism class
of a Calabi-Yau threefold $X$ is classified by its Hodge numbers, $h^{1,1}$ and $h^{2,1}$; its triple intersection numbers, $\kappa_{abc}$; and its second Chern class, $c_2$.

\vspace{.2cm}
\noindent

Positivity of the Hermitian metric on $X$ restricts the K\"ahler form $J$ of $X$, introduced in \eqref{eq:Kahler_form}, to take values in the \textit{K\"ahler cone} $\mathcal{K}_X\subset H^{1,1}(X)\cap H^2(X,\mathbb{R})$, which is equal to the cone of ample line bundles. $\mathcal{K}_X$ is dual to the \textit{Mori cone} of $X$, $\mathcal{M}_X$, which is generated by the effective curve classes of $X$. As effective curves $\mathcal{C}$ are calibrated with respect to the K\"ahler form $J$,
\begin{equation}
	\text{Vol}(\mathcal{C})=\int_\mathcal{C}J\, ,
\end{equation}
it follows that along the facets of $\mathcal{K}_X$, one or more effective curves shrink.

The effective divisor classes in $H^2(X,\mathbb{Z})$ generate the \textit{effective cone}, $\mathcal{E}$, which is dual to the \textit{cone of movable curves} \cite{Boucksom2004}, denoted $\mathrm{Mov}$, generated by the effective curves whose moduli space sweeps out a dense open subset of $X$.

Along some facets of $\mathcal{K}_X$, only effective curves shrink, but no divisors shrink. The curves that can shrink along such facets are always isolated rational curves \cite{atiyah1958analytic,Candelas:1989js}, i.e., $\mathbb{P}^1$'s without continuous moduli spaces, and their normal bundles are isomorphic to $\mathcal{O}(-1)\oplus \mathcal{O}(-1)$, $\mathcal{O}\oplus \mathcal{O}(-2)$, or $\mathcal{O}(1)\oplus \mathcal{O}(-3)$ \cite{Laufer,KatzMorrison}. Continuing past such a facet of $\mathcal{K}_X$, where an effective curve class $[\mathcal{C}]$ shrinks, amounts to a birational morphism
\begin{equation}
	X\backslash \cup_{\mathcal{C}\in [\mathcal{C}]}\mathcal{C}\rightarrow X'\backslash \cup_{\mathcal{C}'\in [\mathcal{C}']}\mathcal{C}'\, ,
\end{equation}
called a flop transition, where $X'$ is in general a topologically distinct Calabi-Yau threefold, and $[\mathcal{C}']$ is an effective curve class in $X'$. Divisors in $X$ are identified with divisors in $X'$ via blowdown maps, and thus we have canonical isomorphisms $H^2(X,\mathbb{Z})\simeq H^2(X',\mathbb{Z})$ and $H_2(X,\mathbb{Z})\simeq H_2(X',\mathbb{Z})$. The Mori cone $\mathcal{M}_{X'}$ differs from $\mathcal{M}_{X}$ because the curve class $[\mathcal{C}]$ ceases to be effective in $X'$, and instead $[\mathcal{C}']\simeq -[\mathcal{C}]$ is effective. Moreover, the Calabi-Yau $X'$ is uniquely characterized (via Wall's theorem) by its intersection numbers and second Chern class
\begin{align}\label{eq:flop_formulas_kappa_c2}
	&\kappa_{abc}'=\kappa_{abc}-\sum_{\mathcal{C}\in [\mathcal{C}]}\mathcal{C}_a \mathcal{C}_b \mathcal{C}_c\, ,\\
	&c'_a=c_a+2\sum_{\mathcal{C}\in [\mathcal{C}]}\mathcal{C}_a\, ,
\end{align}
where $\mathcal{C}_a:=\int_X [\mathcal{C}]\wedge [H_a]$.

Thus, the K\"ahler cones $\mathcal{K}_X$ and $\mathcal{K}_{X'}$ adjoin along a common facet. Exhausting all possible flops in this way, and adjoining their respective K\"ahler cones, one obtains a central object of this work, the \emph{extended K\"ahler cone}
\begin{align}
	\mathcal{K} := \bigcup_{X\in [X]_\text{b}} \mathcal{K}_X \, ,
\end{align}
where $[X]_\text{b}$ is the birational equivalence class of $X$.

Note that some of the cones we have discussed---such as the K\"ahler cone $\mathcal{K}_X$ or the Mori cone $\mathcal{M}_X$---depend on the specific Calabi-Yau threefold $X$, whereas others---such as the extended K\"ahler cone $\mathcal{K}$ and the effective cone $\mathcal{E}$---only depend on the birational equivalence class $[X]_\text{b}$. In our notation, cones in the former class are written with an explicit $X$ subscript, whereas those in the latter class lack such a subscript.

A variety of behaviors are possible at the boundaries of the extended K\"ahler cone. However, as argued in~\cite{Alim:2021vhs}, each known possibility is accompanied by a shrinking (pseudo)effective divisor class. As effective divisors are calibrated with respect to $\frac{1}{2}J\wedge J$, this is best described in \emph{dual coordinates}
\begin{equation}
T_a\df\frac{1}{2}\kappa_{abc}t^b t^c \,,
\end{equation}
which are related to the K\"ahler coordinates by the map
\begin{equation} \label{eq:defcbhtmap}
\mathscr{T}: J \mapsto \frac{1}{2} J \wedge J\,, \qquad \text{i.e.} \qquad  t^a \mapsto \frac{1}{2}\,\kappa_{abc} t^b t^c\,.
\end{equation}
The dual-coordinate image of the extended K\"ahler cone is the \emph{cone of dual coordinates} $\mathcal{T} \df \mathscr{T}(\mathcal{K})$.\footnote{This cone was denoted $\widetilde{\EuScript{K}}$ in \cite{Alim:2021vhs}.} The observation of~\cite{Alim:2021vhs} can then be summarized as $\mathcal{T} = \mathcal{E}^\vee$, i.e., $\mathcal{T}$ is precisely the cone of movable curves.\footnote{It would be interesting to mathematically prove this equivalence, but we view the arguments given in~\cite{Alim:2021vhs} as both compelling and consistent with all known examples.} Since $\mathscr{T}$, viewed as a map from $\mathcal{K}$ to $\mathcal{T}$, is invertible,\footnote{This is proved, e.g., in \S4.1 of \cite{Alim:2021vhs} using the fact that $\mathcal{K}$ is convex.} both $\mathcal{K}$ and $\mathcal{T}$ provide equally good descriptions of the moduli space, with the preferred description depending on the context.

\begin{figure}
	\centering
	\includegraphics[height=6cm]{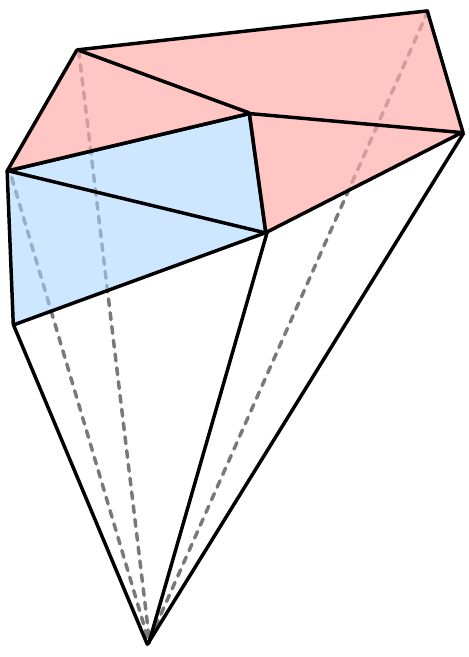}
	\caption{A cartoon of an extended K\"ahler cone. Each cone represents the K\"ahler cone of an individual Calabi-Yau. The cones shaded blue represent Calabi-Yaus obtained as hypersurfaces in a toric variety, whereas those shaded red represent Calabi-Yaus that
are not
manifestly
obtained
by performing
birational transformations on a toric ambient variety, but can nonetheless be reached by flops of curves in the Calabi-Yaus.  One of the main results of this work is a method for assembling extended K\"ahler cones.}
	\label{fig:extended_kahler}
\end{figure}

In practice, a simple way to obtain an ensemble of birationally-equivalent Calabi-Yau threefolds is to consider Calabi-Yau hypersurfaces in simplicial toric fourfolds, defined via distinct fine, regular and star triangulations (FRSTs) of a fixed reflexive polytope $\Delta^\circ$, henceforth referred to as \emph{toric phases}. By adjoining the K\"ahler cones of the different toric fourfolds one obtains a sub-cone of the extended K\"ahler cone $\mathcal{K}$ that includes all birational transformations of the threefold that are inherited from bistellar flips, which are the building blocks of birational transformations of the toric varieties. Henceforth we will refer to such flops of the threefold as \emph{toric flops}. In general, by exploring toric flops one does not obtain all of $\mathcal{K}$, i.e., there generally are flops intrinsic to the threefold for which one lacks an embedding into a pair of birationally-equivalent toric fourfolds. We will denote these as \emph{non-toric flops}. Indeed, by exploring non-toric flops one can
often define new Calabi-Yau threefolds for which one lacks a known hypersurface embedding into any toric variety.
We will refer to such geometric models as \emph{non-toric phases} to distinguish them from hypersurfaces in toric varieties. Assembling the extended K\"ahler cone via collecting and adjoining both toric as well as non-toric phases will be the main focus of this work: see Figure \ref{fig:extended_kahler} for an illustration.

As we will explain in detail in \S\ref{sec:modulispace}, along some facets of $\mathcal{K}$ an effective divisor shrinks to a curve of genus $g$, where the transverse space to each point on the curve is an $A_1$ singularity.
Along such facets, the moduli space ends in a $\mathbb{Z}_2$ orbifold singularity, corresponding to the origin of the Coulomb branch of an $\mathfrak{su}(2)$ gauge sector in the low-energy EFT. One can equally well describe the moduli space redundantly by passing to a covering space of $\mathcal{K}$ that includes its images under a discrete (but not necessarily finite) \emph{Weyl group} $\mathcal{W}$. Although these Weyl images are not genuinely new geometric phases---but rather gauge-redundant copies of the original---this Weyl-extended description is useful not only for describing the BPS black hole solutions~\cite{Alim:2021vhs}, but also (we will find) for the program of moduli space reconstruction.

However, the full Weyl-extended moduli space is in general a \emph{branched cover} of the Weyl orbit of $\mathcal{K}$,
\begin{equation}
\mathcal{W}(\mathcal{K}) \df \bigcup_{w\in\mathcal{W}} w(\mathcal{K}) \,,
\end{equation}
where the branch points arise from CFTs appearing on codimension-two faces of the K\"ahler cone (see the example of \S\ref{app:branch_cuts}). A smaller Weyl extension of moduli space without such subtleties can be obtained by considering only the subgroup $\mathcal{W}_{\text{stable}}$ of Weyl reflections that do not involve crossing any walls of marginal stability, i.e., across which all BPS states remain stable. We refer to such Weyl reflections as \textit{stable} Weyl reflections, and we refer to the orbit of $\mathcal{K}$ under this subgroup as the \emph{hyperextended K\"ahler cone},
\begin{align}\label{eq:defhyp}
\mathcal{K}_{\text{hyp}} := \mathcal{W}_{\text{stable}}(\mathcal{K}) = \bigcup_{w\in \mathcal{W}_{\text{stable}}} w(\mathcal{K})  \,.
\end{align}

Next, we turn to a short discussion of BPS states and their relevant indices, following \cite{Gopakumar:1998jq}. In a compactification of M-theory on $X$, M2-branes wrapped on an effective curve $\mathcal{C}$ give rise to a spectrum of massive BPS particle states that transform under the massive little group $SO(4)\simeq SU(2)_L\times SU(2)_R$ in a representation
\begin{equation}
	R_{[\mathcal{C}]}=\left[\left(\frac{1}{2},0\right)\oplus 2(0,0)\right]\otimes \sum_{j_1,j_2}N^{j_1,j_2}_{[\mathcal{C}]}\cdot (j_1,j_2)\, ,
\end{equation}
with some degeneracies $N^{j_1,j_2}_{[\mathcal{C}]}$. Their contribution to the genus $g$ BPS indices $n^g_{[\mathcal{C}]}$, which are the GV invariants, keeps track of the $SU(2)_L$ representations but traces out the $SU(2)_R$ spin content with sign $(-1)^{2j_2}$. Specifically, the $n_{[\mathcal{C}]}^g$ are computed via the decomposition
\begin{equation}
	\sum_{j_1,j_2} (-1)^{2j_2}(2j_2+1) N^{j_1,j_2}_{[\mathcal{C}]} (j_1)=\sum_{g\geq 0}n^g_{[\mathcal{C}]}\left[\left(\frac{1}{2}\right)+2(0)\right]^g\, ,
\end{equation}
where the right-hand side is a formal sum of vector spaces with possibly negative coefficients. One notes that a multiplet with largest right-spin $j_2$ generically contributes to all $n_{[\mathcal{C}]}^g$ with $g\leq 2j_2$. For instance, hypermultiplets and vector multiplets contribute $+1$ respectively $-2$ to the genus zero GV invariants, and to none of the $n_{[\mathcal{C}]}^{g>0}$.

From a mathematical perspective, the GV invariants $n_{[\mathcal{C}]}^g$ are a particular resummation of Gromov-Witten invariants $\tilde{n}_{[\mathcal{C}]}^g$, which, roughly, count maps from genus $g$ Riemann surfaces into $X$ as computed by the topological A-model \cite{Witten:1988xj,Witten:1991zz}. Specifically, we have the relation \cite{Gopakumar:1998ii,Gopakumar:1998jq,Dedushenko:2014nya}
\begin{equation}
	\sum_{[\mathcal{C}]\in \mathcal{M}_X}\sum_{g=0}^\infty\tilde{n}^g_{[\mathcal{C}]}q^{[\mathcal{C}]}\lambda^{2g-2}=\sum_{[\mathcal{C}]\in \mathcal{M}_X}\sum_{g=0}^\infty\sum_{k=1}^\infty n^g_{[\mathcal{C}]}\frac{1}{k}\Bigl(2\sin\left(\tfrac{k\lambda}{2}\right)\Bigr)^{2g-2}q^{k[\mathcal{C}]}\, ,
\end{equation}
and in particular, at genus zero $\sum_{[\mathcal{C}]\in \mathcal{M}_X}\tilde{n}^0_{[\mathcal{C}]}q^{[\mathcal{C}]}=\sum_{[\mathcal{C}]\in \mathcal{M}_X}n^0_{[\mathcal{C}]}\,\text{Li}_3(q^{[\mathcal{C}]})$, with $\text{Li}_3(q):=\sum_{k=1}^\infty q^k/k^3$.

We will define $\mathcal{M}^{\mathrm{GV}}_{X,g}$ to be the cone generated by all homology classes $[\mathcal{C}]$ of curves with non-vanishing GV invariants $n^g_{[\mathcal{C}]}$. As M2-branes wrapped on curves in classes outside the Mori cone break all the supersymmetries, they do not generate BPS states, and thus their GV invariants $n^g_{[\mathcal{C}]}$ must vanish. We therefore have the inclusion $\mathcal{M}^{\mathrm{GV}}_{X,g}\subseteq \mathcal{M}_X$ for all $g$. The cone $\mathcal{M}^{\mathrm{GV}}_{X}\equiv \mathcal{M}^{\mathrm{GV}}_{X,g=0}$ will be of interest because its generators may collapse at finite distance in moduli space, as we discuss in \S\ref{sec:modulispace}.\footnote{In contrast, the generators of $\mathcal{M}^{\mathrm{GV}}_{X,g>0}$ can never shrink at finite distance in moduli space, as this would lead to finitely many massless higher spin degrees of freedom non-trivially coupled to each other at the two-derivative level, in conflict with classic results \cite{Weinberg:1964,Grisaru:1977}.}

\begin{figure}
	\centering
	\begin{minipage}[t]{0.45\linewidth}
		\includegraphics[keepaspectratio,width=6cm]{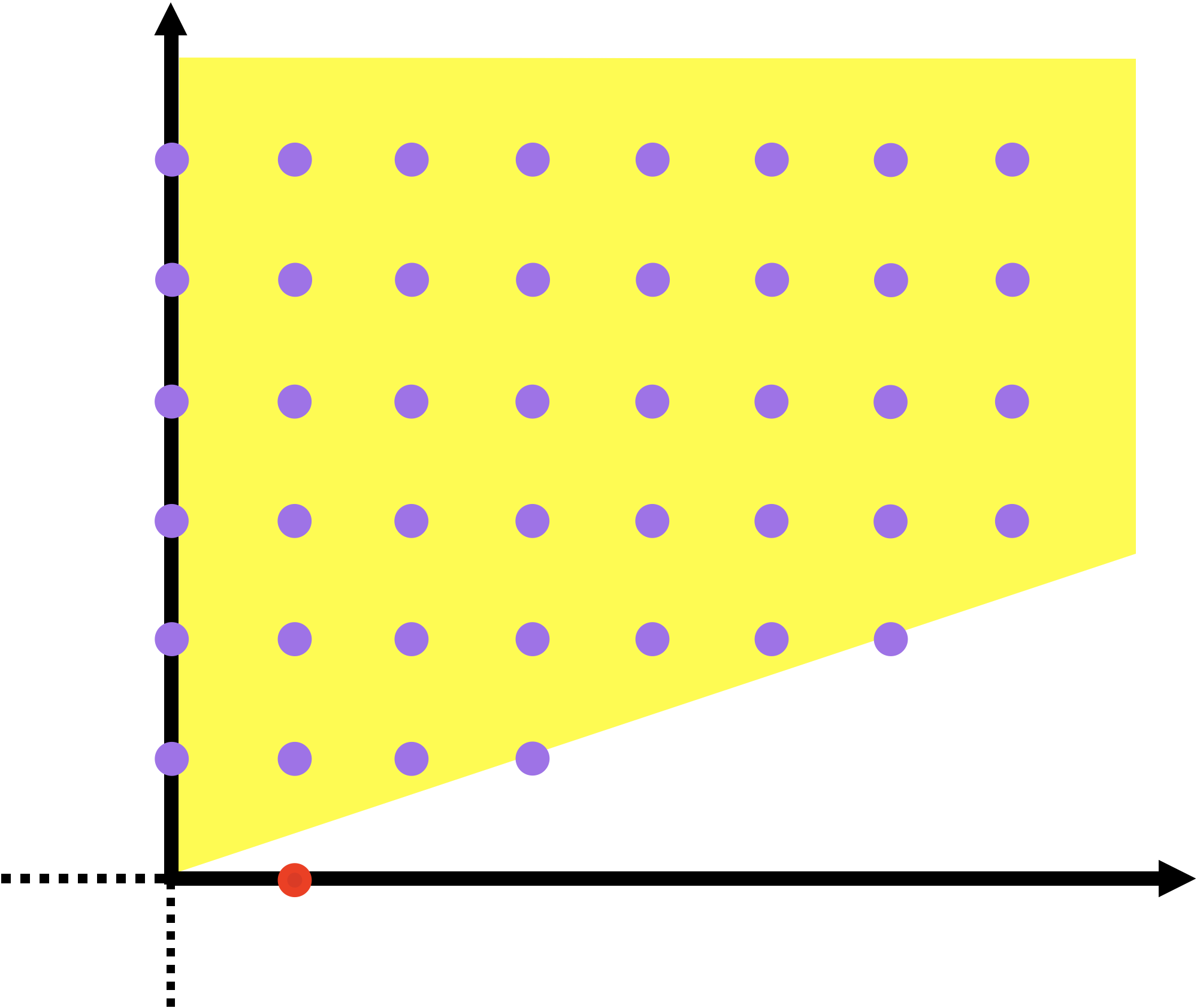}
		\caption{Mori cone $\mathcal{M}_X$ of a Calabi-Yau threefold $X$, and its integer sites populated by non-vanishing genus zero GV invariants. Depicted in red: a generator of $\mathcal{M}_X$ that is nilpotent and lies outside of $\mathcal{M}_\infty$, shown in yellow.}
		\label{fig:GVX}
	\end{minipage}
	\hfil
	\begin{minipage}[t]{0.45\linewidth}
		\centering
		\includegraphics[keepaspectratio,width=6cm]{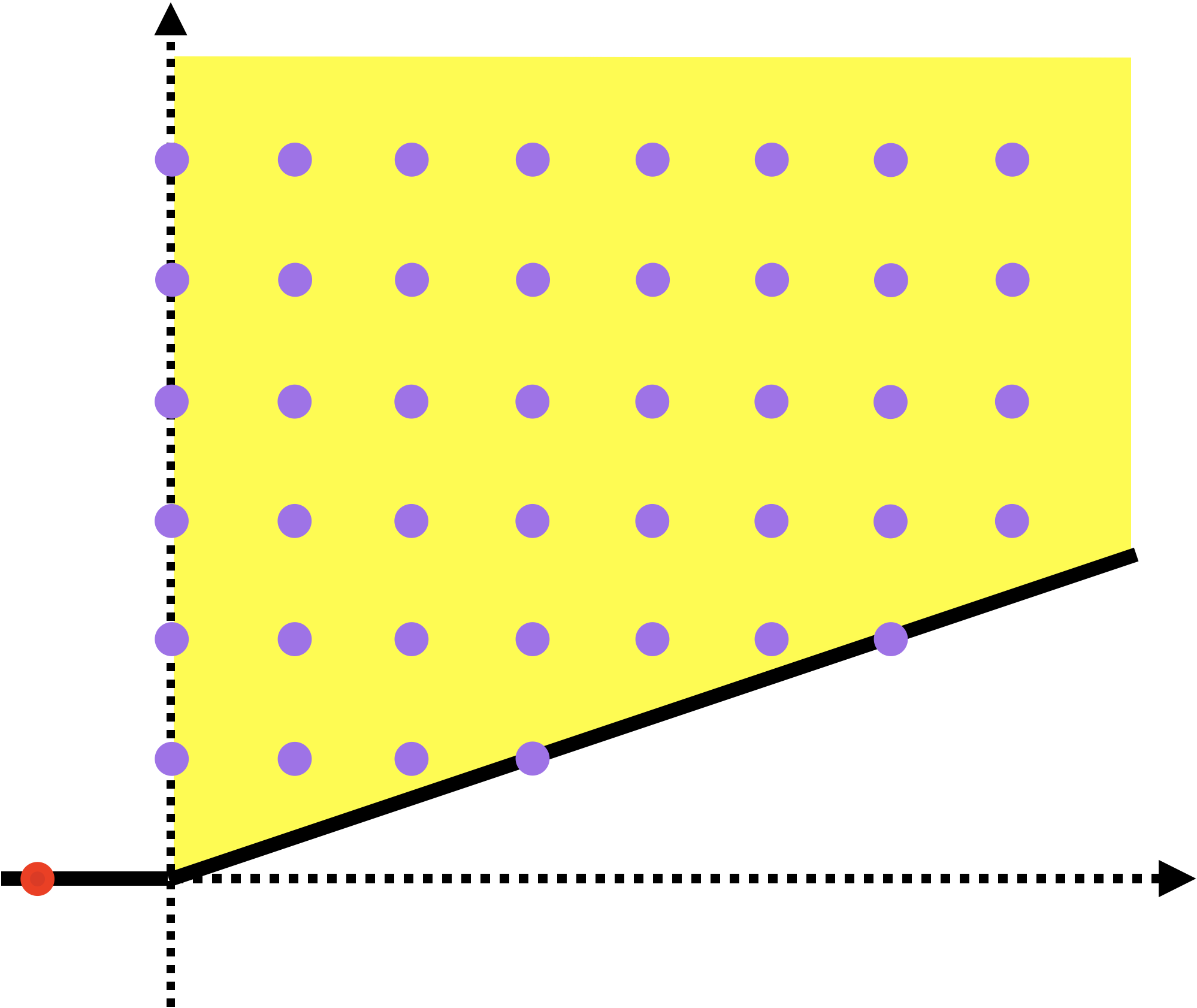}
		\caption{Mori cone $\mathcal{M}_{X'}$ of another Calabi-Yau $X'$ related to $X$ by a flop transition, and sites populated by non-vanishing genus zero GV invariants. Depicted in red: the flopped generator of $\mathcal{M}_X$. In this example, $\mathcal{K}$ is the dual of $\mathcal{M}_\infty$.}
		\label{fig:GVXt}
	\end{minipage}\\
\end{figure}

We will furthermore divide rational rays in $\mathcal{M}_X$ into two classes, following \cite{Demirtas:2020ffz,Demirtas:2021nlu}. In the first class are rays that host an infinite number of curve classes with non-vanishing genus zero GV invariants: these we term \textit{potent rays}.  The second type of ray hosts only a finite number of non-vanishing genus zero GV invariants, and we will call this a \textit{nilpotent ray}, see Figure \ref{fig:GVX} for an example.
Curve classes along potent and nilpotent rays are likewise termed potent and nilpotent, respectively.

An important object in this work is the \textit{infinity cone}, $\mathcal{M}_{\infty}$, defined as the closure of the cone generated by all potent rays.
One of our main claims,  substantiated in \S\ref{sec:modulispace}, is that the dual of the infinity cone is the hyperextended K\"ahler cone $\mathcal{K}_{\text{hyp}}$, i.e.
\begin{equation}
\mathcal{K}_{\text{hyp}} = \mathcal{M}_{\infty}^\vee \,. \label{eq:hypinfty}
\end{equation}

We now relate the aforementioned cones to the charges of BPS black holes. In particular, for any point $T_a^\star$ in the (dual-coordinate) moduli space, let $\mathcal{C}_{\text{BH}}(T_a^\star)$ be the cone of charges of BPS black holes with asymptotic moduli values $T_a \to T_a^\star$.\footnote{Note that $\mathcal{C}_{\text{BH}}(T_a^\star)$ is necessarily convex, because we consider multi-center solutions to be BPS black holes.}
As shown in~\cite{Alim:2021vhs},
\begin{equation}
\mathcal{T} \subseteq \mathcal{C}_{\text{BH}}(T_a^\star) \qquad \text{for any} \qquad T_a^\star \in \mathcal{T} \,.
\end{equation}
This is because $\mathcal{T}$ is convex (since $\mathcal{T} = \mathcal{E}^\vee$) and spherically symmetric BPS attractor flows are straight lines in dual coordinates.

However, since the potent GV invariants remain unchanged across the entire hyperextended moduli space, our checks of the  lattice WGC in fact depend only on the larger (hyperextended) cone of BPS black holes
\begin{equation}
\mathscr{C}_{\text{BH}} \df \bigcup_{T_a^\star \in \mathcal{T}_{\text{hyp}}} \mathcal{C}_{\text{BH}}(T_a^\star) \,,
\end{equation}
where $\mathcal{T}_{\text{hyp}} \df \mathscr{T}(\mathcal{K}_{\text{hyp}})$ is the hyperextended cone of dual coordinates.
In particular, because invariably $T_a^\star \in \mathcal{C}_{\text{BH}}(T_a^\star)$,\footnote{The corresponding BPS black hole solutions are of the five-dimensional Reissner-Nordstr\"om type, with constant values for all the moduli.} we have\footnote{If $\mathcal{T}_{\text{hyp}}$ is convex then the stronger statement $\mathcal{T}_{\text{hyp}} \subseteq \mathcal{C}_{\text{BH}}(T_a^\star)$ also holds. In all examples we have checked, $\mathcal{T}_{\text{hyp}}$ is indeed convex, but we know of no general proof.}
\begin{equation} \label{eq:ThypCBH}
\mathcal{T}_{\text{hyp}} \subseteq \mathscr{C}_{\text{BH}} \,,
\end{equation}
hence the  lattice WGC predicts  BPS particles throughout $\mathcal{T}_{\text{hyp}}$.
One can sometimes establish a stronger inclusion than \eqref{eq:ThypCBH}: see Appendix \ref{sec:CBHcons}.

While the somewhat-complicated relations given above are sufficient for our present purposes, we now argue that the situation may actually be simpler than this. Continuously changing the asymptotic values of the moduli $T_a^\star$  should not change a solution with smooth horizons into a singular solution. Instead, BPS black hole solutions of a given charge may cease to exist another way, by falling apart into infinitely separated components at walls of marginal stability. If these expectations hold, then $\mathcal{C}_{\text{BH}}(T_a^\star)$ can only change at walls of marginal stability. Since by construction there are no such walls within $\mathcal{T}_{\text{hyp}} = \mathscr{T}(\mathcal{K}_{\text{hyp}})$, we would then have
\begin{equation}\label{comeonconjecture}
\mathcal{C}_{\text{BH}}(T_a^\star) \stackrel{?}{=} \mathscr{C}_{\text{BH}} \qquad \text{for any} \qquad T_a^\star \in \mathcal{T}_{\text{hyp}}.
\end{equation}
However, the identification \eqref{comeonconjecture} relies on some non-trivial assumptions that we will not test here, and
in any event the logic of our analysis is valid without \eqref{comeonconjecture}.

By now we have defined a daunting number of cones, some more standard than others: see Table \ref{tab:cones}.
With these definitions in hand, we can give a concise overview of the rest of the paper.
In \S\ref{sec:modulispace}, we will establish the relation $\mathcal{K}_{\text{hyp}} = \mathcal{M}_{\infty}^\vee$ stated in \eqref{eq:hypinfty}.  By computing
$\mathcal{M}_{\infty}$ from the GV invariants of $X$ and applying  \eqref{eq:hypinfty} and \eqref{eq:defcbhtmap}, we arrive at $\mathcal{T}_{\text{hyp}}$, a region inside $\mathscr{C}_{\mathrm{BH}}$.
Finally, by examining the GV invariants of curve classes in $\mathcal{T}_{\text{hyp}}$, we
test the WGC in our ensemble of geometries.

{\renewcommand{\arraystretch}{1.4}
\begin{table}
	\centering
	\begin{tabular}{| c | c | c |}
		\hline
		Symbol & Description & Relations \\
		\hline
		$\mathcal{K}_X$ & K\"ahler cone of phase $X$ & $\equiv\mathcal{M}_X^\vee$\\
		
		$\mathcal{M}_X$ & Mori cone of phase $X$ &  $\equiv\mathcal{K}_X^\vee$ $\vphantom{\bigcap_X}$\\
		
		$\mathcal{K}$ & extended K\"ahler cone & $:= \bigcup_{X \in [X]_{\text{b}}} \mathcal{K}_X $\\
		
		$\mathcal{M}$ & intersection of Mori cones &  $:= \bigcap_{X \in [X]_{\text{b}}} \mathcal{M}_X$, $\equiv\mathcal{K}^\vee $\\
		
		$\mathcal{E}$ & effective cone & $\equiv\text{Mov}^\vee$, $\supseteq \mathcal{K}$ \\
		
		$\text{Mov}$ & cone of movable curves & $\equiv\mathcal{E}^\vee$, $\subseteq \mathcal{K}^\vee$ \\

		$\mathcal{M}_{\infty}$ & infinity cone & $\overset{\eqref{eq:kinfty}}{=} \mathcal{K}^\vee_{\text{hyp}}$  \\

		$\mathcal{M}^{\text{GV}}_{X}$& cone over curves $\mathcal{C}$ with nonzero $n^0_{[\mathcal{C}]}$ & $\overset{\eqref{eq:neq4}}{=}\bigcap_{w\in \mathcal{W}_{\mathcal{N}=4}} w(\mathcal{M}_X)$ \\
		
        $\mathcal{K}_{\text{hyp}}$ & hyperextended K\"ahler cone & $:=\bigcup_{w\in \mathcal{W}_{\text{stable}}} w(\mathcal{K})$, $\overset{\eqref{eq:kinfty}}{=}\mathcal{M}_{\infty}^\vee$ \\

        $\mathcal{T}$ & cone of dual coordinates & $:=\mathscr{T}(\mathcal{K})$, $\stackrel{\!\!\text{\cite{Alim:2021vhs}}}{=}\text{Mov}$ \\

        $\mathcal{T}_{\text{hyp}}$ & hyperextended cone of dual coordinates & $:=\mathscr{T}(\mathcal{K}_{\text{hyp}})$, $\subseteq \mathscr{C}_{\mathrm{BH}}$ \\

        $\mathcal{C}_{\mathrm{BH}}(T_a)$ & cone of BPS black holes at $T_a$ & $ \ni T_a$, $\supseteq \mathcal{T}$\\

        $\mathscr{C}_{\mathrm{BH}}$ & (hyperextended) cone of BPS black holes & $:= \mathcal{C}_{\mathrm{BH}}(\mathcal{T}_{\text{hyp}})$, $\supseteq \mathcal{T}_{\text{hyp}}$\\
		\hline
	\end{tabular}
	\caption{The relevant cones.
The map $\mathscr{T}$ is defined in \eqref{eq:defcbhtmap}. We use $\equiv$ to denote equivalence and $:=$ to denote definitions.
The relations marked $\overset{\eqref{eq:neq4}}{=}$ and $\overset{\eqref{eq:kinfty}}{=}$ are established in the present work, as
\eqref{eq:neq4} and \eqref{eq:kinfty}, respectively, while an argument for the relation marked
$\stackrel{\!\!\text{\cite{Alim:2021vhs}}}{=}$ was given in \cite{Alim:2021vhs}.}
	\label{tab:cones}
\end{table}}

\section{Moduli Space Reconstruction from GV Invariants} \label{sec:modulispace}

We will now explain how the structure of non-vanishing GV invariants determines the hyperextended K\"ahler cone $\mathcal{K}_{\text{hyp}}\supseteq \mathcal{K}$.  We begin in \S\ref{sec:facets} with a review of the classification of possible low energy physics that can arise along facets of the K\"ahler cone $\mathcal{K}_X$ of a Calabi-Yau threefold, paying special attention to the sequence of genus zero GV invariants along the dual generators of the Mori cone. In \S\ref{sec:GV_to_moduli_space} we will use this classification to obtain one of our main results: an algorithm to (1) compute a domain of K\"ahler parameters
$\mathcal{K}_{\text{hyp}}$
that contains the K\"ahler cones of \emph{all} inequivalent Calabi-Yau threefolds $X$ in a given birational equivalence class $[X]_\text{b}$, and (2) compute the defining topological data of all $X\in [X]_\text{b}$.\footnote{See also \cite{Brodie:2020fiq, Brodie:2021ain, Brodie:2021toe, Gendler:2022qof, Lukas:2022crp} for works on applications of flop transitions in Calabi-Yau threefolds.}

\subsection{Facets of the K\"ahler cone}\label{sec:facets}

To reconstruct the extended K\"ahler cone $\mathcal{K}$ starting from a phase $X$, we need to identify every effective curve class $[\mathcal{C}]$ that can be flopped to reach a new phase $X'$, and then adjoin to $\mathcal{K}_{X}$ the associated phase $\mathcal{K}_{X'}$.
To this end, we now review, following Witten \cite{Witten:1996qb}, the possible low energy physics that can arise along facets of the K\"ahler cone $\mathcal{K}_X$, and what imprint the corresponding light charged states leave on the spectrum of GV invariants.  This will allow us to use knowledge of the set of nonzero GV invariants to
differentiate curves that can be flopped from those that shrink on exterior boundaries of $\mathcal{K}$.

\subsubsection{Flop transitions}
Flop transitions always occur at finite distance in moduli space, and the light states that arise on the corresponding facets of $\mathcal{K}_X$ are furnished by M2-branes wrapping the isolated volume-minimizing representatives of the shrinking curve class $[\mathcal{C}]$, i.e.~a set of $\mathbb{P}^1$'s, contributing a finite number of massless hypermultiplets \cite{Strominger:1995cz}. The number of such multiplets is equal to the genus zero GV invariant of the vanishing curve class. Furthermore, at a generic point on such a facet, at most finitely many states fall within any finite mass window, and thus the class $[\mathcal{C}]$ is nilpotent (as defined in \S\ref{sec:cygeometry}) and lies strictly outside of the cone $\mathcal{M}_{\infty}$.

\subsubsection{Weyl reflections} \label{sec:weyl}

Next we consider a finite distance facet along which an effective curve $\mathcal{C}$, again a $\mathbb{P}^1$, shrinks to a point in $X$, while simultaneously an effective divisor $D$ degenerates to a genus $g$ Riemann surface $\mathcal{R}$ worth of $A_1$ singularities. In this case $D$ is a $\mathbb{P}^1$ fibration over $\mathcal{R}$, where the fiber is the shrinking $\mathbb{P}^1$. Along such facets of $\mathcal{K}_X$, a $\mathfrak{u}(1)$ factor in the generic gauge algebra enhances to $\mathfrak{su}(2)$, and one obtains $g$ charged hypermultiplets \cite{Aspinwall:1995xy,Katz:1996ht}.\footnote{As explained in \cite{Alim:2021vhs}, along such facets the generic gauge group $U(1)^n$ enhances to either $SU(2)\times U(1)^{n-1}$, or $U(2)\times U(1)^{n-2}$, or $SO(3)\times U(1)^{n-1}$.} These massless electrically charged states come from M2-branes wrapped on the generic fiber, while the M5-brane wrapped on the vanishing divisor is interpreted as the 't Hooft-Polyakov magnetic monopole string. The genus zero GV invariant of $[\mathcal{C}]$ is equal to $2g-2$. If the $\mathbb{P}^1$ fibration over $\mathcal{R}$ defining the shrinking divisor $D$ degenerates over $N_F$ points in the base, one obtains in addition $N_F$ fundamentally charged hypermultiplets from $2N_F$ isolated rigid $\mathbb{P}^1$'s in half the class of the generic fiber (see \S\ref{app:fundamentally_charged} for an example), and if $N_F>0$ the gauge group is $SU(2)$.

In flat coordinates (as opposed to gauge-invariant coordinates) it makes sense to continue past the facet where $[\mathcal{C}]$ and $[D]$ shrink, but the corresponding region in moduli space is gauge equivalent to the original K\"ahler cone via the Weyl group action of the non-abelian gauge theory \cite{Alim:2021vhs},
\begin{equation}\label{eq:Weyl-reflection}
	t^a\simeq {w^a}_b\cdot t^b\, ,\quad {w^a}_b={\delta^a}_b-2\frac{[D]^a[\mathcal{C}]_b}{\langle \mathcal{C},D \rangle}\, .
\end{equation}
In general, we define the Weyl group $\mathcal{W}$ associated to a birational equivalence class $[X]_\text{b}$ as the group generated by all such Weyl reflections across facets of the extended K\"ahler cone $\mathcal{K}$.

The shrinking curve class $[\mathcal{C}]$ is either a generator of $\mathcal{M}_\infty$ or else lies strictly outside of $\mathcal{M}_\infty$. To understand this, we first consider the $g=0$
case without fundamentally charged matter, where the non-abelian gauge theory in question is pure supersymmetric Yang-Mills theory.  The four-dimensional theory is asymptotically free, and its Coulomb branch receives corrections from gauge instantons that dominate over the classical result at Coulomb branch vevs smaller than the dynamical scale of the theory, as famously computed by Seiberg and Witten \cite{Seiberg:1994rs}. These corrections can only be accounted for by the known expression for the prepotential \eqref{eq:prepotentialIIA} if there are non-vanishing genus zero GV invariants along an \emph{infinite} sequence of curves that grow by multiples of the vanishing curve, i.e.~if $n^{0}_{[\mathcal{C}^\prime_k]}\neq 0$ for sequences of curve classes of the form
\begin{equation}\label{eq:affine_curve_ray}
	[\mathcal{C}^\prime_k] := [\mathcal{C}^\prime_0]+k\cdot [\mathcal{C}]\, ,\quad k=0,\ldots,\infty\, ,
\end{equation}
or infinite sub-sequences thereof. This implies that $[\mathcal{C}]$ is a generator of $\mathcal{M}_\infty$, even though the GV sequence associated with its own integer multiples terminates, i.e.~even though $[\mathcal{C}]$ itself is nilpotent. More generally, whenever the non-abelian gauge theory is asymptotically free, i.e.~if~$4(1-~g)-N_F>0$, the shrinking class is a generator of $\mathcal{M}_\infty$.

\begin{figure}
	\centering
	\begin{minipage}[t]{0.45\linewidth}
		\includegraphics[keepaspectratio,width=6cm]{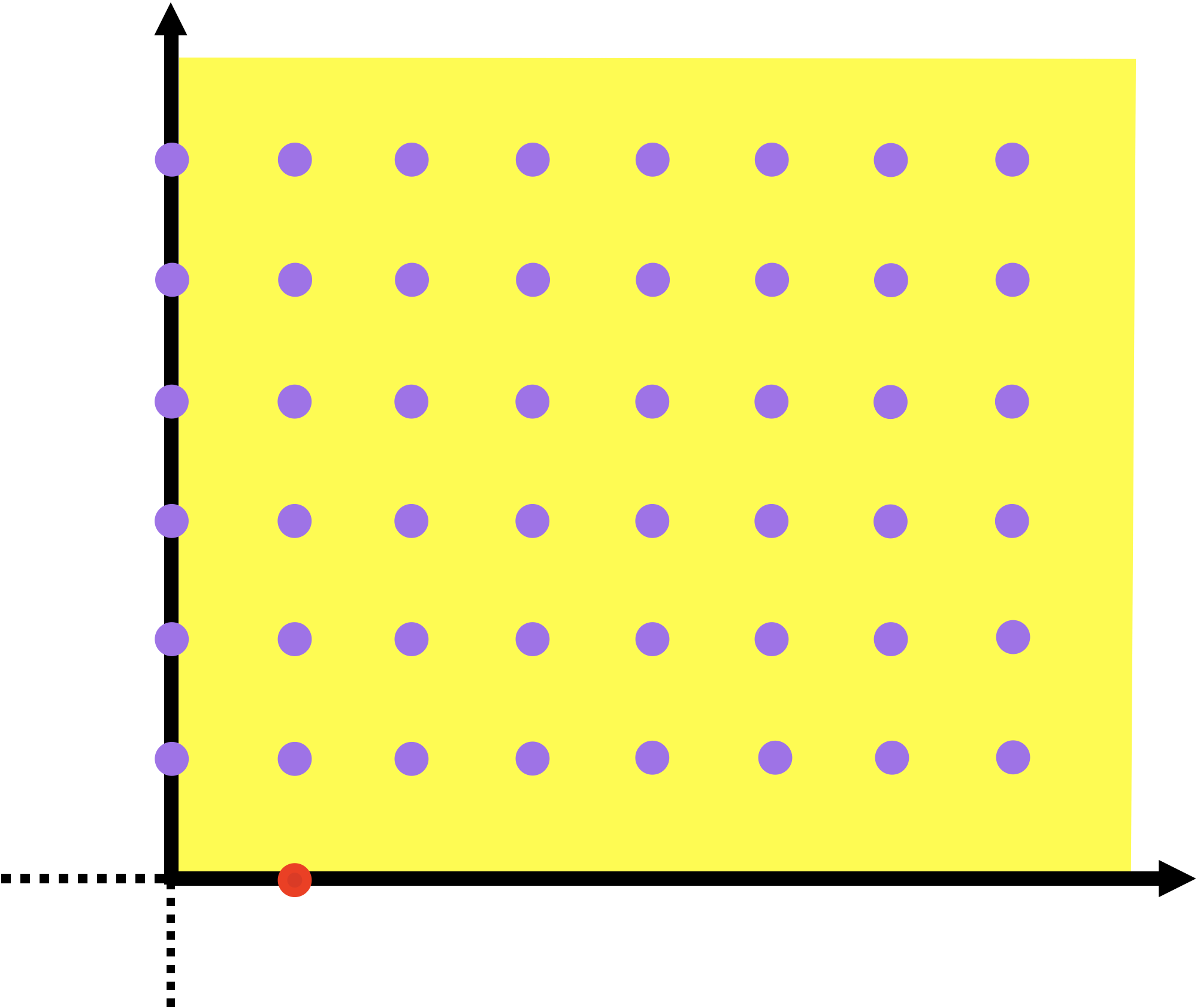}
		\caption{Mori cone $\mathcal{M}_X$ of a Calabi-Yau threefold $X$, and its integer sites populated by non-vanishing genus zero GV invariants. Depicted in red: a generator of $\mathcal{M}_X$ that is nilpotent and lies on the boundary of $\mathcal{M}_{\infty}$.}
		\label{fig:boundaryflop}
	\end{minipage}
	\hfil
	\begin{minipage}[t]{0.45\linewidth}
		\centering
		\includegraphics[keepaspectratio,width=6cm]{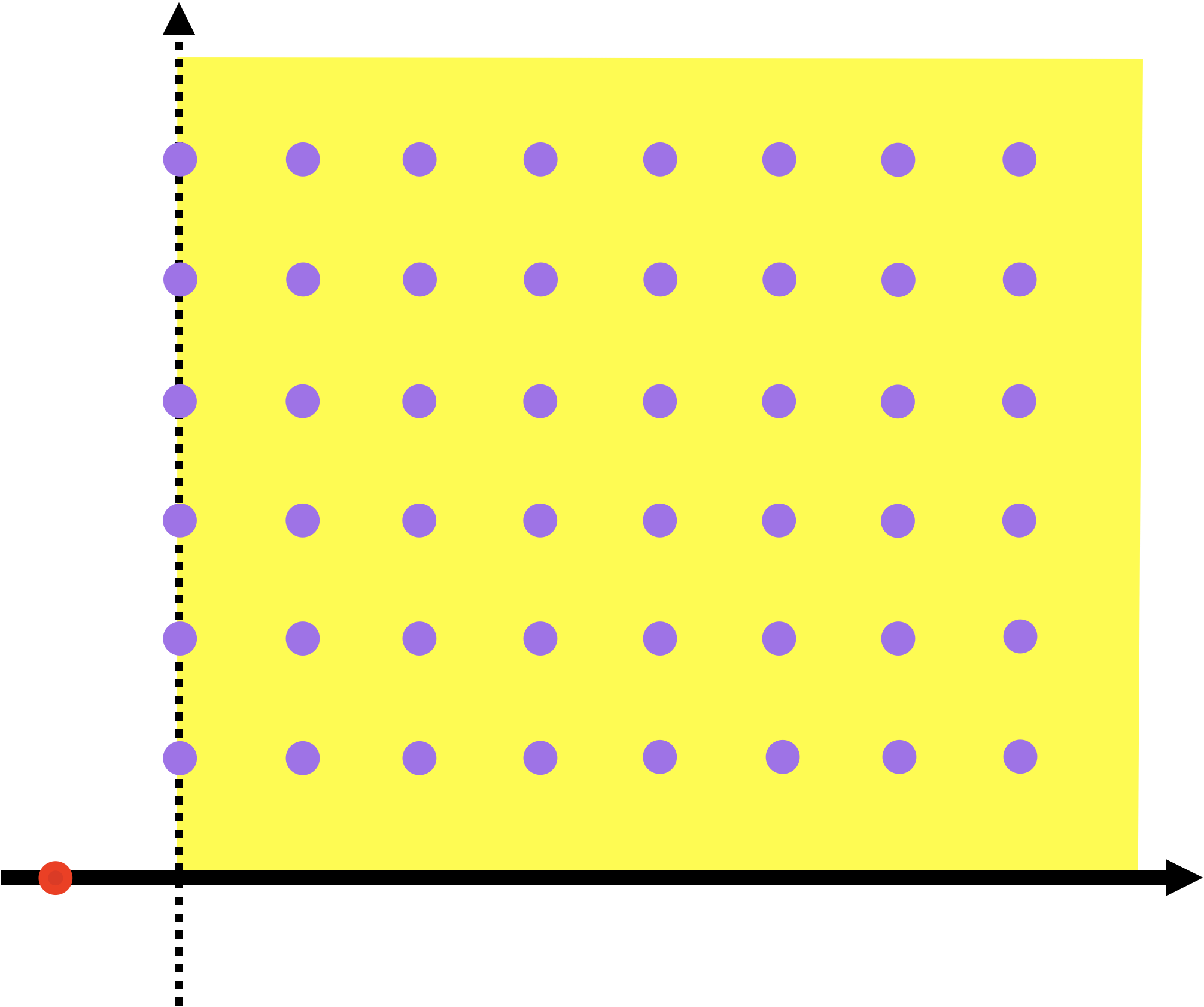}
		\caption{Non-pointed Mori cone $\mathcal{M}_{X'}$ related to $\mathcal{M}_{X}$ by an unstable Weyl flop, if one assumes no wall-crossing. Depicted in red: the flopped generator of $\mathcal{M}_X$.}
		\label{fig:boundaryflopped}
	\end{minipage}
\end{figure}

As an aside we note that this implies the existence of a wall of marginal stability already in the five-dimensional theory at the origin of the Coulomb branch: the fact that there exists at least one sequence of the form \eqref{eq:affine_curve_ray} implies that upon continuing into a distinct Weyl chamber --- by passing through the origin of the Coulomb branch --- all but finitely many BPS indices associated to curve classes in the series must jump to zero. Otherwise, the Weyl-transformed Mori cone could not be a pointed cone, as illustrated in Figures~\ref{fig:boundaryflop} and \ref{fig:boundaryflopped}.

The converse statement is false: gauge instanton corrections to the prepotential of an IR-free gauge theory vanish at the origin of the Coulomb branch, but one may still have sequences of the form \eqref{eq:affine_curve_ray} that give rise to a holomorphic series with a finite analytic continuation to the origin of the Coulomb branch (see \S\ref{app:fundamentally_charged} for an example). Thus, a curve $\mathcal{C}$ shrinking along a Weyl-reflection locus associated with non-abelian enhancement giving rise to an IR-free gauge theory in four dimensions can be either a generator of $\mathcal{M}_{\infty}$ or lie strictly outside of it.

As far as the spectrum of GV invariants is concerned, a facet of $\mathcal{K}_X$ featuring non-abelian enhancement with $g>1$ hypermultiplets in the adjoint representation and no fundamentally charged hypermultiplets cannot be distinguished from a flop of $2g-2$ rational curves.\footnote{Similarly, a genus $g>1$ non-abelian enhancement with $N_F$ fundamentally charged hypermultiplets cannot be distinguished from a length-2 flop with GV invariants $(2N_F,2g-2)$.} Indeed, Weyl flops of genus $g>0$ occur only along tuned loci in complex structure moduli space, and turn into ordinary flops (if $g>1$) or smooth points (if $g=1$) under a generic perturbation in complex structure \cite{wilson-92}.
We will single out the case where $g=1$, again without fundamental matter: here the GV invariant of the shrinking curve vanishes, due to the fact that the non-abelian gauge theory is conformal and has accidentally enhanced $\mathcal{N}=4$ supersymmetry.  When such a $g=1$ locus exists, the cone over non-vanishing genus zero GV invariants $\mathcal{M}_X^{\text{GV}}$ is strictly smaller than the Mori cone $\mathcal{M}_X$, and the cone dual to $\mathcal{M}_X^{\text{GV}}$ is the K\"ahler cone adjoined by its gauge equivalent copy generated by the Weyl reflection associated with the $\mathcal{N}=4$ Yang-Mills theory.\footnote{We assume that no walls of marginal stability can arise at the origin of the Coulomb branch of a gauge theory with enhanced supersymmetry.}  We denote the group generated by such Weyl reflections as $\mathcal{W}_{\mathcal{N}=4}$.

To recap, there are three kinds of qualitatively different Weyl reflections that can occur, distinguished by the sign of the beta function in the four-dimensional gauge theory.
\begin{itemize}
	\item For an asymptotically free gauge theory, i.e.~$4(1-g)-N_F>0$, there is an infinite tower of gauge instanton corrections that become important near the origin of the Coulomb branch of the four-dimensional theory. As a consequence, the shrinking generator of the Mori cone is also a generator of $\mathcal{M}_\infty$.
	\item In the conformal case $4(1-g)-N_F=0$, the shrinking generator $[\mathcal{C}]$ of the Mori cone is either strictly outside of $\mathcal{M}_\infty$ or is on its boundary. In the special case $(g,N_F)=(1,0)$ the gauge theory enhances to the $\mathcal{N}=4$ theory, and we expect that $[\mathcal{C}]$ is strictly outside $\mathcal{M}_\infty$.
	\item For an infrared-free gauge theory, i.e.~$4(1-g)-N_F<0$, the shrinking generator of the Mori cone can be a generator of $\mathcal{M}_\infty$ or lie strictly outside of $\mathcal{M}_\infty$.
\end{itemize}
In general, $\mathfrak{su}(2)$ enhancements from generators of $\mathcal{M}_\infty$ have a wall of marginal stability at their respective origins of the Coulomb branch (corresponding to unstable Weyl reflections), while those from curves outside of $\mathcal{M}_\infty$ do not (corresponding to stable Weyl reflections).

\subsubsection{Tensionless string CFTs} \label{tcft}
A third possibility is that  an effective curve $[\mathcal{C}]$ shrinks, while simultaneously an effective divisor degenerates to a point in $X$. As explained in \cite{Witten:1996qb}, an infinite tower of M2-brane excitations on curves in $D$ becomes massless, while the magnetically charged string from the M5-brane wrapped on $D$ becomes tensionless. One arrives at a non-trivial SCFT in five dimensions featuring  tensionless strings \cite{Witten:1996qb}.  The presence of infinitely many electrically charged BPS states --- barring infinitely many exact cancellations in the index, which we will assume does not happen --- implies that multiples of the shrinking curve class $[\mathcal{C}]$ itself contribute an infinite sequence of nonzero GV invariants. Such a point lies at finite distance in moduli space.  One might expect that a point where an infinite number of states becomes massless is necessarily at infinite distance \cite{Grimm:2018ohb, Heidenreich:2018kpg}, but in this case,
strong coupling in the infrared invalidates this intuition.

\subsubsection{Asymptotic boundaries}\label{asymp}
The final possibility is a facet of the K\"ahler cone where a curve class $[\mathcal{C}]$ and effective divisor class $[D]$ shrink at infinite distance in moduli space. This corresponds to a partial decompactification of the five-dimensional theory, and for reasons as in \S\ref{tcft} one expects that the GV sequence $n^0_{k[\mathcal{C}]}$ does not terminate.\footnote{In the infinite distance limit, this infinite tower of nonvanishing GV invariants leads to an infinite tower of light particles, in agreement with the Distance Conjecture of Ooguri and Vafa \cite{Ooguri:2006in}.
Such limits have been considered in e.g.~\cite{Grimm:2018cpv,Corvilain:2018lgw,Gendler:2020dfp}.}

\subsection{Moduli space from genus zero GV invariants}\label{sec:GV_to_moduli_space}

We will now apply the classification given in \S\ref{sec:facets} to see how flop curves can be identified from their GV invariants.

If $[\mathcal{C}]$ is a potent curve, i.e.~if the GV sequence $n^0_{k[\mathcal{C}]}$ does not terminate, then $[\mathcal{C}]$ cannot be flopped: shrinking $[\mathcal{C}]$ corresponds at best to a facet of the K\"ahler cone that is either a tensionless string CFT or an asymptotic boundary.

Next, if $[\mathcal{C}]$ is a nilpotent curve in $\mathcal{M}_{\infty}$ --- either strictly inside or on the boundary --- that is not a generator of $\mathcal{M}_{\infty}$,
then one cannot shrink $[\mathcal{C}]$ without first shrinking a generator.  Thus without loss of generality we can consider curves that are generators of $\mathcal{M}_{\infty}$.
A nilpotent generator of $\mathcal{M}_{\infty}$ is not a flop curve, but instead corresponds to an unstable Weyl reflection.

The only remaining case is that $[\mathcal{C}]$ is nilpotent and lies strictly outside the cone $\mathcal{M}_{\infty}$ of potent curves.
We call such a curve a nilpotent-outside-potent, or \emph{nop}, curve.  The key lesson of \S\ref{sec:facets} is that
\begin{align}
[\mathcal{C}]\text{ is a nop curve } \Leftrightarrow \text{ shrinking $[\mathcal{C}]$ yields a flop or a stable Weyl reflection\,,}
\end{align}
where we refer to a Weyl reflection as stable if it results from $\mathfrak{su}(2)$ enhancement \emph{without} wall-crossing at the origin of the Coulomb branch.
One important consequence is that by identifying all nop curves, we determine the group $\mathcal{W}_{\text{stable}} \subseteq \mathcal{W}$ of stable Weyl reflections.

Analytically continuing past a wall in the K\"ahler cone where a nop curve shrinks leads either to a topologically distinct Calabi-Yau, or to a Calabi-Yau that gives rise to the same physics as the original geometry, either due to a gauge symmetry or an isomorphism of the two geometries.

The discussion above leads us to several useful results. First, the cone over the non-vanishing genus zero GV invariants $\mathcal{M}_X^{\text{GV}}$ is equal to the intersection over the
$\mathcal{N}=4$ Weyl orbit of the Mori cone:
\begin{equation}\label{eq:neq4}
	\mathcal{M}_X^{\text{GV}}=\bigcap_{w\in \mathcal{W}_{\mathcal{N}=4}} w(\mathcal{M}_X)\, .
\end{equation}
As $\mathcal{W}_{\mathcal{N}=4}$ is non-trivial only for tuned complex structure, we further have that $\mathcal{M}_X^{\text{GV}}=\mathcal{M}_X$ at generic points in complex structure moduli space.

Second, the cone over the rays in the Mori cone hosting infinite series of GV invariants, which we denoted $\mathcal{M}_{\infty}$, is equal to the intersection of all Mori cones and their images under the stable Weyl group, $\mathcal{W}_{\text{stable}}$:
\begin{equation}
	\mathcal{M}_\infty=\bigcap_{w\in \mathcal{W}_{\text{stable}}}\bigcap_{X\in [X]_\text{b}} w(\mathcal{M}_X)\equiv \bigcap_{w\in \mathcal{W}_{\text{stable}}} w(\mathcal{M}) \, .
\end{equation}
In particular, the dual   of $\mathcal{M}_{\infty}$ is the union over the $\mathcal{W}_{\text{stable}}$ orbit of the extended K\"ahler cone,
\begin{equation}\label{eq:kinfty}
	\mathcal{M}_{\infty}^\vee=\bigcup_{w\in \mathcal{W}_{\text{stable}}}w(\mathcal{K}) =: \mathcal{K}_{\text{hyp}}\,,
\end{equation}
which is what we previously defined as the hyperextended K\"ahler cone $\mathcal{K}_{\text{hyp}}$.
We have therefore established the relation \eqref{eq:hypinfty}. For generic complex structure we have $\mathcal{K}_{\text{hyp}}=\mathcal{K}$.

In summary, one can determine the full K\"ahler moduli space of M-theory compactifications from the data of genus zero GV invariants, up to possible overcounting of gauge-equivalent chambers.

\subsection{Geometric data from analytic continuation} \label{sec:geometricdata}
We have explained how to determine an overcomplete domain for K\"ahler moduli, $\mathcal{K}_{\text{hyp}}:=\mathcal{W}_{\text{stable}}(\mathcal{K})$, furnished by the extended K\"ahler cone and its orbit under stable Weyl reflections.
But, in addition to charting out the domain of K\"ahler moduli one would also like to characterize the Calabi-Yau threefold, and evaluate the prepotential $\mathcal{F}^{\text{IIA}}$, in each geometric chamber. It turns out that computing the four-dimensional prepotential $\mathcal{F}^{\text{IIA}}$ in one particular representative $X$ of a birational equivalence class $[X]_\text{b}$ is sufficient to determine the defining geometric data (and thus the large volume expansion of the prepotential) in \textit{all} chambers of $\mathcal{K}_{\text{hyp}}$, in a way that is readily accessible.

We will now consider how the prepotential \eqref{eq:prepotentialIIA} transforms across a facet on which a nop curve shrinks.  In this scenario, the shrinking curve formally acquires negative volume on the other side of the transition. Along the
facet where the nop curve has zero volume, the volumes of all curves in $\mathcal{M}_\infty$ can be made arbitrarily large by scaling up the K\"ahler parameter, and for each choice of K\"ahler parameter there exist at most finitely many curves in $\mathcal{M}_\infty$ that have volumes below any fixed finite threshold. Thus, in order to obtain the geometric data of all different chambers, one must only make sense out of formally negative volumes appearing in finitely many polylogarithms in the expression \eqref{eq:prepotentialIIA}.\footnote{The reasoning here is formally similar to that employed in \cite{Demirtas:2020ffz} to compute Calabi-Yau periods systematically near conifold boundaries of the large complex structure cone. Indeed, for a flop curve the limit of vanishing curve volume is mirror dual to the conifold limit discussed in \cite{Demirtas:2020ffz}.} This is straightforward to do by invoking Jonqui\`ere's identity
\begin{align}
	\frac{\mathrm{Li}_3(e^{2\pi i u})}{(2\pi i)^3} = \frac{\mathrm{Li}_3(e^{2\pi i (-u)})}{(2\pi i)^3} -\frac{u^3}{6} + \frac{u^2}{4} - \frac{u}{12}\, ,
\end{align}
which allows one to rewrite an instanton correction with formally negative action as one with positive action, supplemented by a polynomial correction term. Geometrically, this is interpreted as collapsing a holomorphic curve class $[\mathcal{C}]$, continuing past its vanishing locus, and re-interpreting its formally negative volume as minus the volume of the new effective curve class $[\mathcal{C}']=-[\mathcal{C}]$. The polynomial correction term is interpreted as the perturbative correction induced by the passing through the locus where a hypermultiplet has become massless, and can be absorbed into the classical geometric data, precisely reproducing \eqref{eq:flop_formulas_kappa_c2} for the case of a flop, where the genus zero GV invariant is a literal count of shrinking curves, while also generalizing to the case of a stable Weyl reflection.  In summary, upon passing through a flop transition, or a Weyl reflection without wall-crossing, whereby an effective nop curve $[\mathcal{C}]$ collapses, the classical geometric data is modified as
\begin{align}\label{eq:flop_formulas_kappa_c2'}
	&\kappa_{abc}'=\kappa_{abc}-n^0_{\mathcal{C}}\,\mathcal{C}_a \mathcal{C}_b \mathcal{C}_c\, , \nonumber\\
	&c'_a=c_a+2n^0_{\mathcal{C}}\,\mathcal{C}_a\, ,
\end{align}
and the GV invariants remain the same, up to reassigning
\begin{equation}\label{eq:flop_formulas_GV}
	n'^0_{[\mathcal{C}']}\equiv n'^0_{-[\mathcal{C}]}= n^0_{[\mathcal{C}]}\, ,\quad \text{and}\quad n'^0_{[\mathcal{C}]}=0\, .
\end{equation}

\subsection{Assembly algorithm} \label{sec:algorithm}

We can now state an algorithm for constructing the hyperextended K\"ahler cone $\mathcal{K}_{\text{hyp}}$ of a Calabi-Yau threefold $X$:
\begin{enumerate}
\item Compute the genus-zero GV invariants of $X$, up to high enough degree to obtain an accurate approximation to $\mathcal{M}_{\infty}$.
\item Identify all nop curves, i.e.~nilpotent curves on extremal rays that are strictly outside $\mathcal{M}_{\infty}$.
\item Assemble $\mathcal{K}_{\text{hyp}}$ by adjoining phases related by flops and Weyl reflections, out to the boundaries of the moduli space listed in \S\ref{tcft} and \S\ref{asymp}.
\end{enumerate}
In practice, some of the nilpotent curves in step (2) may correspond to stable Weyl reflections, so that the corresponding chambers do not furnish physically new regions of moduli space.
In fact, the number of chambers of $\mathcal{W}_{\text{stable}}(\mathcal{K})$ is often quite large, or even infinite, as a consequence of
symmetries.  We explain in Appendix \S\ref{sec:symmetries} how we replace $\mathcal{K}_{\text{hyp}}$ with a smaller, less redundant cone, containing a finite number of chambers. Using this, our assembly algorithm is applicable even to birational equivalence classes featuring infinitely many symmetric flops, and non-polyhedral effective cones such as the examples of \cite{Brodie:2021ain,Gendler:2022qof}. Analogously, our algorithm can be applied to cases where the effective cone becomes non-polyhedral at special loci in complex structure moduli space, corresponding to an infinite order $\mathcal{W}_{\text{stable}}$.

In contrast, if $\mathcal{M}_{\infty}$ is infinitely generated due to the existence of infinitely many flops that do not arise as simple reflections across facets of K\"ahler cones, then any computation of GV invariants at finite cutoff degree will yield only an (inner) approximation of $\mathcal{M}_{\infty}$, which asymptotes to the exact result as the cutoff degree is sent to infinity. Similarly, if any representative in a birational equivalence class  has a non-polyhedral Mori cone, then GV invariants computed to finite cutoff degree will produce only an approximation of the true Mori cone, and likewise $\mathcal{M}_{\infty}$. A famous such example is the Schoen manifold \cite{Schoen}.

\subsection{An example}\label{sec:example}

As an example, let us consider a Calabi-Yau threefold hypersurface $X$ with $h^{1,1}=2$ favorably embedded in a toric fourfold $V$, constructed in a standard way\footnote{For some details on Calabi-Yau threefold hypersurfaces in toric fourfolds, see Appendix \ref{app:ToricTech}.  Further examples are given in Appendix \ref{app:examples}.} from a reflexive polytope $\Delta^\circ$. The integer points in $\Delta^\circ$ other than the origin are the columns of
\begin{equation}
	\begin{pmatrix}
		-1&  0&  0&  0& -1&  1\\
		-1&  0&  0&  1&  1&  0\\
		-1&  0&  1&  0&  0&  0\\
		-1&  1&  0&  0&  0&  0
	\end{pmatrix}\, ,
\end{equation}
and are associated with homogeneous coordinates $x_I$, $I=1,\ldots,6$, identified via a $(\mathbb{C}^*)^2$ action with scaling weights
\begin{align}\label{eq:glsm_main_example}
\begin{bmatrix}
x_1 & x_2 & x_3 & x_4 & x_5 & x_6\\ \hline
0 & 0 & 0 & -1 & 1 & 1\\
1 & 1 & 1 & 2 & -1 & 0
\end{bmatrix}\, .
\end{align}
The columns of \eqref{eq:glsm_main_example} are the charges of the divisors $D_I:=\{x_I=0\}\cap X$. The faces of $\Delta^\circ$ are simplices, so $V$ is smooth.

In a basis of divisor classes $\{[D_1],[D_6]\}$, the triple intersection numbers and second Chern class of $X$ are
\begin{equation}
\kappa_{1ab}=\begin{pmatrix}
5 & 5 \\
5 & 5
\end{pmatrix}\, ,\quad \kappa_{2ab}=\begin{pmatrix}
5 & 5 \\
5 & 3
\end{pmatrix}\, ,\quad c_a=\begin{pmatrix}
50\\
42
\end{pmatrix}\, .
\end{equation}
The Mori cone inherited from the ambient variety, $\mathcal{M}_V$, is generated by the curve classes $(1, 0)$ and $(0, 1)$.
The genus zero GV invariants are given in Table \ref{GVextab_main},
\begin{table}
	\scriptsize \renewcommand{\arraystretch}{0.9}
	\begin{align*}
		\begin{array}{c|cccccccccc}
			\mathdiagbox[width=0.7cm,height=0.5cm,innerleftsep=0.1cm,innerrightsep=0cm]{q_1}{q_2} & 0 & 1 & 2 & 3 & 4 & 5 \\ \hline
			0 &           \cellcolor{pink}    *&        \cellcolor{yellow}      56&    \cellcolor{yellow}        -272&  \cellcolor{yellow}          3240&   \cellcolor{yellow}       -58432&     \cellcolor{yellow}    1303840\\
			1& 20&         \cellcolor{pink}   2635&     \cellcolor{pink}       2760&   \cellcolor{yellow}       -45440&  \cellcolor{yellow}       1001340& \cellcolor{yellow}      -26330880\\
			2 & 0&         \cellcolor{yellow}   5040&      \cellcolor{pink}    541930&   \cellcolor{pink}       933760&   \cellcolor{pink}    -18770880&   \cellcolor{yellow}    490600080\\
			3& 0&        \cellcolor{yellow}     190&   \cellcolor{yellow}      2973660&   \cellcolor{pink}    277421695&   \cellcolor{pink}    563282580&  \cellcolor{pink}  -11813767700\\
			4& 0&          \cellcolor{yellow}   -40&    \cellcolor{yellow}     2454600&   \cellcolor{yellow}   2644224240&    \cellcolor{pink}208000930200&   \cellcolor{pink} 470459159880\\
			5 & 0&       \cellcolor{yellow}        3&     \cellcolor{yellow}      67980&  \cellcolor{yellow}    5829698942&  \cellcolor{yellow} 2855250958116& \cellcolor{pink} 193028959075965\\
			6& 0&               0&       \cellcolor{yellow}   -14960& \cellcolor{yellow}     3084577280& \cellcolor{yellow} 11119027471400&\cellcolor{yellow}3465883673329200\\
			7& 0&               0&        \cellcolor{yellow}    3420&   \cellcolor{yellow}     75341270& \cellcolor{yellow} 14592676836440&\cellcolor{yellow}19950547779012810\\
			8&0&               0&      \cellcolor{yellow}      -760&   \cellcolor{yellow}    -13884400&  \cellcolor{yellow} 5711374027440&\cellcolor{yellow}45586693863580200\\
			9&0&               0&       \cellcolor{yellow}      100&   \cellcolor{yellow}      2767590&   \cellcolor{yellow} 132960571500&\cellcolor{yellow}42020108300555745\\
			10&0&               0&        \cellcolor{yellow}      -6&      \cellcolor{yellow}   -783664&  \cellcolor{yellow}  -21741657848&\cellcolor{yellow}13122339863069280
		\end{array}
	\end{align*}
	\caption{Genus zero GV invariants $n_{q_1, q_2}^0$ for the geometry of \S\ref{sec:example}. Charges inside the infinity cone $\mathcal{M}_\infty = \left\{ q_1-5q_2, q_2 \geq 0 \right\}$ but outside the cone of dual coordinates $\mathcal{T} = \left\{  q_2-q_1, 2q_1 -q_2 \geq 0 \right\}$ are shown in yellow. Charges in $\mathcal{T}$ are shown in pink.}
	\label{GVextab_main}
\end{table}
and the sites in $\mathcal{M}_V$ populated by them are depicted in Figure \ref{fig:GVX}. In particular, the generators of $\mathcal{M}_V$ have non-vanishing GV invariants, so $\mathcal{M}_V \simeq \mathcal{M}_X$, and the generator class $[\mathcal{C}_1]:=(1,0)$ is nilpotent.\footnote{Computing the exact Mori cone by showing that the GV invariants of the torically inherited cone $\mathcal{M}_V$ are non-zero has also been employed in earlier works, e.g., \cite{Anderson:2017aux,Gendler:2022qof}.}  Moreover, the infinity cone $\mathcal{M}_{\infty}$ is generated by the classes $(5,1)$ and $(0,1)$. Upon shrinking the curve class $[\mathcal{C}_2]:=(0,1)$ the divisor $D_5$ shrinks to a point, generating a tensionless string CFT. Indeed, the sequence of GV invariants along this generator of the Mori cone is infinite:
\begin{equation}
	n^0_{k\cdot[\mathcal{C}_2]}=\{56, -272, 3240, -58432, 1303840, -33255216,\ldots\}
\end{equation}
We now note that the curve class $[\mathcal{C}_1]$ lies outside $\mathcal{M}_{\infty}$ (and it is the only curve class that does). Therefore, one can flop this curve\footnote{One can check that no divisors shrink in this limit, and so this is a true flop.}
\begin{equation}
[\mathcal{C}_1]\longrightarrow -[\mathcal{C}_1]\, ,
\end{equation}
obtaining a smooth threefold $X'$ with genus zero GV invariants as depicted in Figure \ref{fig:GVXt}. This flop is not inherited from any birational transformation of the toric ambient variety, so is non-toric in the sense described in \S\ref{sec:cygeometry}.

As this phase is non-toric, we will determine its Mori cone from the data of GV invariants. The generators of $\mathcal{M}_{X'}^{\text{GV}}=\cap_{w\in \mathcal{W}_{\mathcal{N}=4}}w(\mathcal{M}_{X'})$ are $-[\mathcal{C}_1]=(-1,0)$ and $[\mathcal{C}_3]:=(5,1)$. The flopped triple intersection numbers and second Chern class, expressed in their natural basis, the generators $(-[\mathcal{C}_1],[\mathcal{C}_3])$ of $\mathcal{M}_{X'}^{\text{GV}}$, are
\begin{equation}
	\kappa'_{1ab}=\begin{pmatrix}
		90 & 30\\
		30 & 10
	\end{pmatrix}\, ,\quad \kappa'_{2ab}=\begin{pmatrix}
	30 & 10\\
	10 & 3
\end{pmatrix}\, ,\quad c_a'=\begin{pmatrix}
120\\
42
\end{pmatrix}\, ,
\end{equation}
showing that $X'$ is not isomorphic to $X$.

If $\mathcal{W}_{\mathcal{N}=4}$ were non-trivial, it would have to map the set of generators of $\mathcal{M}_{X'}^{\text{GV}}$ to themselves. Thus, in this example, at least one element would have to exchange its two generators. However, this cannot be a symmetry as the GV invariants of the two generators of $\mathcal{M}_{X'}^{\text{GV}}$ are distinct: they turn out to be $20$ and $3$ respectively (cf. Table \ref{GVextab_main}). Therefore $\mathcal{W}_{\mathcal{N}=4}$ is trivial and we can determine the exact Mori cone: $\mathcal{M}_{X'}=\mathcal{M}_{X'}^{\text{GV}}$. Along the facet of $\mathcal{K}_{X'}$ where the curve class $[\mathcal{C}_3]$ shrinks, the toric divisor $D_4$ shrinks to a point, leading to another tensionless string CFT. Again, the sequence of GV invariants along this generator of the Mori cone $\mathcal{M}_{X'}$ is infinite:
\begin{equation}
	n^0_{k\cdot[\mathcal{C}_3]}=\{3, -6, 27, -192, 1695, -17064,\ldots\}
\end{equation}

The extended K\"ahler cone $\mathcal{K}$ is equal to the dual of the infinity cone $\mathcal{M}_\infty$ in this example, and, expressed in the original basis of divisor classes $([D_1],[D_6])$,  it is generated by the classes $(1,0)$ and $(-1,5)$. Absent any Weyl group we have $\mathcal{K}_{\text{hyp}}=\mathcal{K}$. The cone of dual coordinates $\mathcal{T}$ is generated by $(1,1)$ and $(1,2)$ and is indeed dual to the effective cone, generated by the divisors $[D_4]=(-1,2)$ and $[D_5]=(1,-1)$.

\section{Checks of the Weak Gravity Conjecture} \label{sec:wgc}

The WGC \cite{Arkani-Hamed:2006emk} provides an interesting constraint on the spectra of effective theories that arise as low-energy limits of quantum gravity.
We will now describe a direct check of the WGC in a large ensemble of Calabi-Yau hypersurface
compactifications of M-theory.  This analysis relies on using
Gopakumar-Vafa invariants to
reconstruct the K\"ahler moduli spaces of Calabi-Yau threefolds, via the algorithm we described in \S\ref{sec:modulispace}.

\subsection{The lattice WGC for BPS states}
The WGC states that in an effective field theory that admits a UV-completion to a theory of quantum gravity, black holes must be able to decay.  In its simplest form, for a theory with a single $U(1)$ gauge field, this condition states there must be a (super)extremal particle, i.e., a particle whose charge to mass ratio satisfies
\begin{align}
\biggl|\frac{Q}{M}\biggr| \geq \biggl|\frac{Q}{M}\biggr|_{\text{ext}},
\label{WGCeq}
\end{align}
where the quantity on the right-hand side is the charge to mass ratio of an extremal black hole.

In theories with more than one $U(1)$ gauge field, the condition that black holes must be able to decay translates to the condition that the convex hull of charge to mass ratios of particles in the spectrum must contain the region in charge to mass space where black holes can exist \cite{Cheung:2014vva}. The convex hull condition is the minimal requirement for black holes to decay in a given theory, but a more stringent version of the conjecture was proposed \cite{Heidenreich:2016aqi} for reasons of consistency under dimensional reduction (see also \cite{Montero:2016tif, Andriolo:2018lvp}).
This stronger version, known as the lattice WGC, holds that \emph{every} site in the charge lattice must support a (super)extremal particle.

In full generality, the lattice WGC is difficult to check: the spectrum of light states is unknown, and for theories with moduli, the black hole extremality bound deviates non-trivially from the Reissner-Nordstr\"om result. Nonetheless, in certain regions of charge space, one can take advantage of the properties of BPS states.
Recall that we have denoted by $\mathscr{C}_{\mathrm{BH}}$ the cone in charge space where BPS black holes exist.
The lattice WGC is satisfied in the BPS region of charge space if and only if for each site in $\mathscr{C}_{\mathrm{BH}}$, there exists a (super)extremal state.

Following \cite{Alim:2021vhs}, in the region $\mathscr{C}_{\mathrm{BH}}$ the BPS bound implies that a (super)extremal state is indeed extremal, and BPS. Thus, the lattice WGC implies the existence of a single-particle BPS state for every charge in $\mathscr{C}_{\mathrm{BH}}$.
We will apply this fact to test the lattice WGC by using
GV invariants to count BPS states in a subcone of $\mathscr{C}_{\mathrm{BH}}$.
Specifically, as explained in \S\ref{sec:cygeometry}, we have the containment relation $\mathcal{T}_{\text{hyp}}\subseteq \mathscr{C}_{\mathrm{BH}}$, where $\mathcal{T}_{\text{hyp}}$ is the cone of dual coordinates defined in \eqref{eq:defcbhtmap}. In \S\ref{sec:GV_to_moduli_space} we established how to compute $\mathcal{T}_{\text{hyp}}$ from the data of GV invariants, and thus we are now equipped to test the lattice WGC in $\mathcal{T}_{\text{hyp}}$.

\subsection{Testing the lattice WGC}\label{sec:wgtest}

We have applied the above logic to check the lattice WGC in a large ensemble of compactifications of M-theory on Calabi-Yau threefolds constructed as hypersurfaces in toric varieties.  We used {\tt{CYTools}} \cite{Demirtas:2022hqf,computational-mirror-symmetry} to compute genus-zero GV invariants, and so carried out the algorithm outlined in \S\ref{sec:algorithm}.

In fact, in this setting one can first perform a more efficient test of a sufficient condition for the lattice WGC, and for many geometries this test obviates the full process of \S\ref{sec:algorithm}.  The process is as follows: one finds the cone $\mathcal{E}_{V}$ of effective divisors inherited from the ambient toric variety $V$, which obeys $\mathcal{E}_V\subseteq \mathcal{E}$, and therefore
\begin{equation}\label{eq:inclu}
\mathcal{T}_{\text{hyp}} \subseteq \mathcal{E}_{V}^\vee\,.
\end{equation}
The analogue of \eqref{eq:inclu} also holds
if $\mathcal{E}_{V}$ is replaced by any other inner bound on the true effective cone $\mathcal{E}$.
A sufficient test of the lattice WGC is to check that every site in $\mathcal{E}_{V}^\vee$ is populated with a non-vanishing GV invariant.
If there are empty sites in $\mathcal{E}_{V}^\vee$, then either the lattice WGC is false, or there exist  \textit{autochthonous} divisors, i.e.~effective divisors in the Calabi-Yau that are not inherited from the ambient variety. In this way, the lattice WGC makes non-trivial predictions for geometry, specifically for the cone of effective divisors (cf.~\cite{Demirtas:2019lfi,Long:2021lon}).

A subclass of autochthonous divisors that we will refer to as \emph{min-face} divisors are readily obtained from polytope data, by examining simple factorizations of the defining polynomial.  In practice, we construct the cone $\mathcal{E}_{\text{inner}}$ generated by all inherited effective divisors, as well as any min-face divisors, which obeys
\begin{equation}
 \mathcal{E}_{V} \subseteq \mathcal{E}_{\text{inner}} \subseteq \mathcal{E}\,.
\end{equation}
We then check whether there are empty sites in $\mathcal{E}_{\text{inner}}^\vee$, up to some cutoff degree.  If there are no such empty sites, the lattice WGC holds in the example in question, up to the cutoff degree tested.

In cases where $\mathcal{E}_{\text{inner}}^\vee$ contains empty sites, then either the lattice WGC is false, or there are further autochthonous divisors that are not of min-face type.  To analyze such cases, we proceed by using the GV invariants to obtain  $\mathcal{T}_{\text{hyp}} \subseteq \mathscr{C}_{\mathrm{BH}}$, as described in \S\ref{sec:geometricdata}.  This approach is more computationally expensive, because one has to adjoin all phases resulting from flops, but it is also exhaustive.

We can now describe the general algorithm to check the lattice WGC in the BPS cone
of a Calabi-Yau threefold, $X$.  Given an arbitrarily chosen grading vector $\vec{v}_g\in H^2(X,\mathbb{Z})$ that lies strictly interior to $\mathcal{K}$, we define the degree of a charge $\vec{q}\in H_2(X,\mathbb{Z})$ as  $d=\vec{v}_g \cdot \vec{q}$. Then, up to a cutoff degree $d \leq d_{\text{cutoff}}$, we implement the following procedure:
\begin{enumerate}
\item For each site $q\in \mathcal{E}_{\text{inner}}^\vee$ with $d_q\leq d_{\text{cutoff}}$, compute the GV invariant. If all sites have non-vanishing GV invariants, the lattice WGC is satisfied up to cutoff degree $d_{\text{cutoff}}$. If not, continue.
\item Using the methods described in \S\ref{sec:geometricdata}, identify all possible flops  of $X$, and calculate triple intersection numbers in each one.
\item Compute $\mathcal{T}_{\text{hyp}}$ using \eqref{eq:kinfty} and \eqref{eq:defcbhtmap}.
\item For each site in $q\in \mathcal{T}_{\text{hyp}}$ with degree less than $d_{\text{cutoff}}$, compute the GV invariant. If all such sites have non-vanishing GV invariant, the lattice WGC passes this test.
\end{enumerate}

Because this approach involves a computation up to a finite cutoff degree, there are several subtleties associated to possible misidentifications.
For example, an apparent violation of the lattice WGC could in principle arise due to an overestimation of $\mathscr{C}_{\text{BH}}$, though in all examples considered in this paper we have been able to compute GV invariants to high enough order to avoid this issue. In \S\ref{sec:appendixfinite} we give a comprehensive discussion of the potential problems of a finite computation, and we explain how we have mitigated them.

\begin{figure}[h!]
\centering
  \begin{subfigure}[t]{.45\linewidth}
    \centering\includegraphics[width=\linewidth]{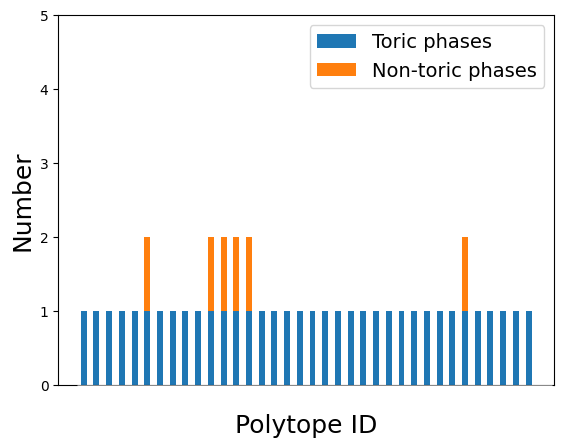}
    \caption{$h^{1,1}=2$}
  \end{subfigure}
  \begin{subfigure}[t]{.45\linewidth}
    \centering\includegraphics[width=\linewidth]{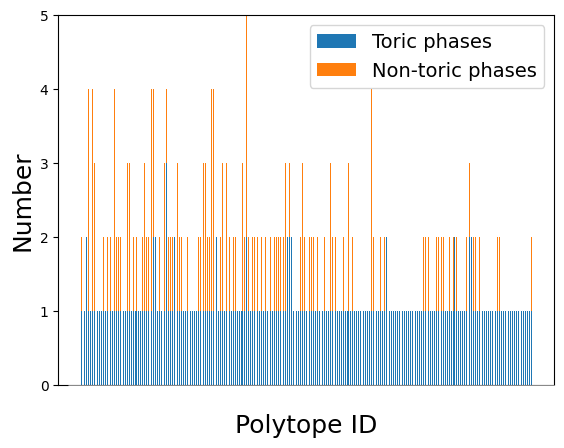}
    \caption{$h^{1,1}=3$}
  \end{subfigure}\\
  \centering
  \begin{subfigure}[t]{.45\linewidth}
    \centering\includegraphics[width=\linewidth]{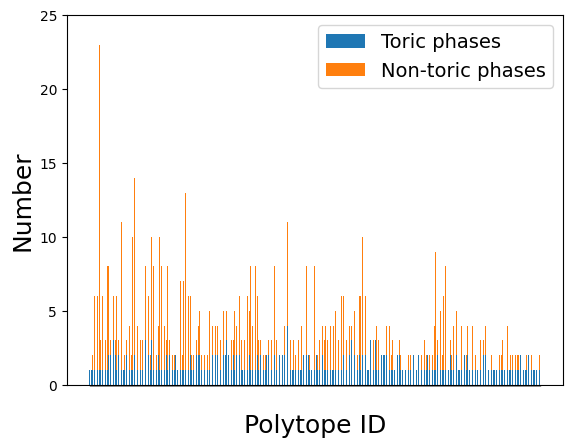}
    \caption{$h^{1,1}=4$}
  \end{subfigure}
  \caption{The number of apparently-inequivalent toric and non-toric phases that we found.  For $h^{1,1}=2,3,4$ we have (36,\,6), (274,\,123), and (1760,\,2180) toric and non-toric phases, respectively.  The polytope IDs are ordered as in \cite{Kreuzer:2000xy}, and in particular $h^{2,1}$ increases to the right.}
  \label{fig:toric_v_nontoric}
 \end{figure}

We have performed the above analysis beginning with a set of seed Calabi-Yau threefolds, from which we obtain the rest of the phases via flop transitions.
The $2062$  seed threefolds result from a single FRST of each of the $1464$ four-dimensional reflexive polytopes with $2 \le h^{1,1} \le 4$, as well as
$598$ additional favorable polytopes with $h^{1,1}=5$.
For each seed geometry $X$, we compute the GV invariants and subsequently identify the geometric phases of the given birational equivalence class $[X]_{\text{b}}$, obtained via flop transitions from the original phase $X$.

One of the key results of this work is that, via flop transitions of toric hypersurface Calabi-Yaus, we can discover Calabi-Yau threefolds that cannot be written as hypersurfaces in the initial ambient toric variety, as in the example presented in \S\ref{sec:example}. In fact, this method sometimes allows one to discover Calabi-Yaus that cannot be obtained via a triangulation of \textit{any} four-dimensional reflexive polytope.\footnote{This is the case for the non-toric phase in the example in \S\ref{sec:example}.} But in general we do not distinguish between the two cases in this work, and we use `non-toric phase' to refer to a phase for which we have found no manifest description as a hypersurface in a toric variety.
Figure \ref{fig:toric_v_nontoric} shows the  number of geometric phases obtained from FRSTs of  polytopes, as well as the number of non-toric phases obtained through flops.

We carried out the algorithm outlined above to test the lattice WGC in $2062$ polytopes.  We found no violations of the lattice WGC.

It should be noted that there are known counterexamples to the lattice WGC, all of which arise from compactifications on toroidal orbifolds \cite{Heidenreich:2016aqi} (see also the recent work \cite{Montero:2022vva}).  Nonetheless, in the present work we have found no evidence against the proposition that the lattice WGC holds in M-theory compactifications on Calabi-Yau threefold hypersurfaces.
Even so, we hasten to remind the reader that the present computation is in a limited ensemble, at $h^{1,1} \le 5$, and only checks the lattice WGC in the BPS sector, up to a finite cutoff degree.

We also note that in $1170$ of these cases the lattice WGC predicted autochthonous divisors, and of these cases, $493$ geometries predicted further autochthonous divisors beyond min-face ones.  These predictions were then confirmed with the full algorithm.

\section{Conclusions}\label{sec:conclusions}

A primary result of this work is a method for computing the complete K\"ahler moduli space of a Calabi-Yau threefold $X$ from knowledge of its genus zero Gopakumar-Vafa invariants.

We have shown that an efficient way to identify all phase transitions in K\"ahler moduli space is to compute $\mathcal{M}_{\infty} \subset H_{2}(X)$, i.e.~the closure of the cone generated by rays supporting infinite series of nonvanishing GV invariants. Curves in the complement of $\mathcal{M}_{\infty}$ within the Mori cone shrink across phase transitions. Some of these are flop transitions, while others are Weyl reflections associated with non-abelian gauge enhancements. By passing through all such phase transitions, unto the boundaries of moduli space, one assembles an object that we call the hyperextended K\"ahler cone, $\mathcal{K}_{\text{hyp}}$. In the absence of Weyl reflections, $\mathcal{K}_{\text{hyp}}$ coincides with the extended K\"ahler cone $\mathcal{K}$, while in general $\mathcal{K}_{\text{hyp}} \supseteq \mathcal{K}$ gives a gauge-redundant parametrization of $\mathcal{K}$.

We carried out this algorithm in an ensemble of geometries, including all favorable hypersurfaces with $1 \le h^{1,1} \le 4$  resulting from the Kreuzer-Skarke list, as well as 598 examples with $h^{1,1}=5$.  Using {\tt{CYTools}} \cite{Demirtas:2022hqf} to compute GV invariants, we constructed the K\"ahler moduli spaces in each case.

The other major result of this work is a test of the Weak Gravity Conjecture in compactifications of M-theory on Calabi-Yau threefold hypersurfaces.
The lattice WGC predicts that in the region in charge space where BPS black holes exist, which we denoted $\mathscr{C}_{\mathrm{BH}}$, there must exist a complete lattice of BPS particles.
We showed that a subcone $\mathcal{T}_{\text{hyp}}$ of $\mathscr{C}_{\mathrm{BH}}$ is determined by the data of $\mathcal{K}_{\text{hyp}}$: see~\eqref{eq:ThypCBH}.
Moreover, a nonzero GV invariant for a given charge implies the existence of a corresponding BPS particle.
Thus, GV invariants serve a dual purpose: they determine $\mathcal{M}_{\infty}$, from which we obtain $\mathcal{K}_{\text{hyp}}$ and in turn the region $\mathcal{T}_{\text{hyp}}$ where we test the WGC, and they also provide information on the BPS particle spectrum, which is the quantity being tested.

In the geometries we studied, we checked the prediction of the WGC for every charge in $\mathcal{T}_{\text{hyp}} \subseteq \mathscr{C}_{\mathrm{BH}}$ up to a finite cutoff degree.  We found that not just the tower and sublattice WGC, but even the full lattice WGC, holds for the charges in question.
We stress that our computation provides no information about BPS particles beyond the chosen cutoff degree, nor any information about non-BPS particles.
Thus, our result should be read as confirming the predictions of the lattice WGC in a finite subregion of the charge lattice.

\section*{Acknowledgments}

We thank Manki Kim, Nate MacFadden, Jake McNamara, Richard Nally, Andres Rios-Tascon, Andreas Schachner, and Mike Stillman for helpful discussions.
N.G., L.M., and J.M. were supported in part by NSF grant PHY-1719877. B.H. was supported by NSF grants PHY-1914934 and PHY-2112800.
T.R. was supported in part by the Berkeley Center for Theoretical Physics; by the Department of Energy, Office of Science, Office of High Energy Physics under QuantISED Award DE-SC0019380 and under contract DE-AC02-05CH11231; and by the National Science Foundation under Award Number 2112880.

\appendix
\addtocontents{toc}{\protect\setcounter{tocdepth}{1}}%

\section{Computational Algorithm}
In this Appendix, we give details of the algorithm for reconstructing the K\"ahler moduli space and testing the WGC.  We discuss the residual gauge redundancy, as well as the possible misidentifications of flops that can result from a computation of the GV invariants of finitely many curves.

\subsection{Identifying flop curves and stable Weyl reflections} \label{sec:nops}

We first describe how to identify flop curves given a Calabi-Yau obtained as a hypersurface in a toric variety, together with the charges and associated GV invariants computed up to a finite cutoff degree $d_{\mathrm{cutoff}}$ using a grading vector $\vec{v}_g$ (given by a suitable point in the K\"ahler cone). We will denote by $\mathfrak{C}$ the set of charges with nonzero GV invariants up to this cutoff.

We start by creating a list $\mathfrak{N}$ of
curve classes that appear nilpotent based on a computation of GV invariants up to cutoff degree $d_{\mathrm{cutoff}}$.
To do this, for each co-prime curve class $[\mathcal{C}]$ we check if multiples of $[\mathcal{C}]$ have nonzero GV invariants, up to the cutoff $d_{\mathrm{cutoff}}$. Given a charge $[\mathcal{C}]$, if there exists a positive integer multiple $k^*$ such that
\begin{align}
	k^* [\mathcal{C}] \cdot \vec{v}_g \leq d_{\mathrm{cutoff}}\qquad \text{and} \qquad  n_{k^*[\mathcal{C}]}^0 = 0\, ,
\end{align}
and $\sum_{k=1}^{k^*-1}k^2 \cdot n^0_{k[\mathcal{C}]}>0$,
then the charge $[\mathcal{C}]$ is added to $\mathfrak{N}$.

Given the set of (apparently) nilpotent charges $\mathfrak{N}$ and the set of (apparently) potent charges $\mathfrak{C} \backslash \mathfrak{N}$, we would like to determine the set  $ \mathfrak{F} \subseteq \mathfrak{N}$ of nop curves. This is done in two steps:
\begin{enumerate}
	\item For each curve $[\mathcal{C}] \in \mathfrak{N}$, determine whether $[\mathcal{C}]$ is on the boundary of the closure of the cone generated by $\mathfrak{C} \backslash \mathfrak{N}$. Intuitively, if $[\mathcal{C}]$ lies strictly outside of the cone generated by $\mathfrak{C} \backslash \mathfrak{N}$, the ray that passes through $[\mathcal{C}]$ diverges from all potent rays (and even other nilpotent rays). In practice we will test for this feature in the following non-unique way: we find the largest integer $k$ such that $k[\mathcal{C}]$ lies on or below the affine subspace defined by \emph{half} the cutoff degree, i.e. $k[\mathcal{C}]\cdot \vec{v}_g\leq d_{\mathrm{cutoff}}/2$. Then, having determined $k$, we define a codimension one lattice by the intersection of the affine subspace generated by all curves $[\mathcal{C}']$ with degree equal to $k[\mathcal{C}]\cdot \vec{v}_g$, declaring $k[\mathcal{C}]$ as the new origin, and find a reduced lattice basis using LLL reduction \cite{lenstra-lll}. In the lattice-reduced basis we compute $d$, defined as the smallest norm $||\cdot ||_\infty$ of all integer points that arise from rays in $\mathfrak{C} \backslash \mathfrak{N}$. We use this quantity as an integer notion of distance between the candidate nilpotent ray and the potent rays along a slice through the Mori cone defined by half the cutoff degree.
	Then, we repeat this process, similarly computing an integer distance $d'$ but with $d_{\text{cutoff}}/2$ replaced by the full cutoff degree $d_{\text{cutoff}}$. If $d'>d$, we conclude that, up to the chosen cutoff degree, the ray passing through $[\mathcal{C}]$ diverges from the potent rays and is thus a candidate nop curve. We add it to a list of candidate nop curves, $\mathfrak{F}_0$.
	\item Assuming that all curves strictly interior to $\mathcal{M}_\infty$ with sufficiently large degree are in fact potent, step (1) will reliably identify all nop curves of sufficiently low degree. However, one frequently finds examples where regions inside $\mathcal{M}_\infty$ are not populated by potent rays (see e.g. the GV invariants in Table \ref{GVextab2} of our example \S\ref{app:fundamentally_charged}). In order to deal with this we repeat step (1) for all the candidate nop curves in $\mathfrak{F}_0$ with the following difference: in computing the pair $(d,d')$ we take the norm of \emph{all} integer points that arise from rays in the convex cone generated by the complement $\mathfrak{C} \backslash \mathfrak{F}_0$, as opposed to just the potent rays.
	All curves in $\mathfrak{F}_0$ that still appear to diverge from the potent rays via this notion of distance are added to the list $ \mathfrak{F}$, and treated as nop curves.
\end{enumerate}
It is not hard to convince oneself that for every nop curve there exists a finite cutoff degree such that the above algorithm will find the nop curve. Similarly, every nilpotent ray on the boundary of $\mathcal{M}_\infty$ will be rejected for sufficiently large cutoff. Conversely, for any fixed cutoff $d_{\text{cutoff}}$ one may over or underestimate the true set of nop curves, a subtlety which we discuss in more detail in \S\ref{sec:appendixfinite}.

\subsection{Gauge equivalence and isomorphisms} \label{sec:symmetries}

The set $\mathfrak{F}$ is the set of curves that can be flopped in some phase to reconstruct the hyperextended K\"ahler cone, $\mathcal{K}_{\mathrm{hyp}}$. Many (sometimes infinitely many---see \cite{Gendler:2022qof}) chambers of $\mathcal{K}_{\mathrm{hyp}}$ are redundant, either because of gauge symmetries or topological isomorphisms between the Calabi-Yaus. In practice, we construct a less redundant subset of the hyperextended K\"ahler cone, $\mathcal{K}_{\circ} \subseteq \mathcal{K}_{\mathrm{hyp}}$, by identifying the curves in $\mathfrak{F}$ that, when flopped, produce Weyl-symmetric or isomorphic phases. To this end, given a particular Calabi-Yau $X$ and the generators of $\mathcal{M}^{\text{GV}}_X$, we perform the following test:

\begin{enumerate}
	\item For each $\mathcal{M}^{\text{GV}}_X$ generator $[\mathcal{C}] \in \mathfrak{F}$, one tests whether there exists a non-vanishing $\vec{d}\in H^{1,1}(X,\mathbb{Z})$, such that $\kappa_{abc}d^a t^b t^c\equiv 0$ along the dual facet of $\mathcal{K}_X$. If not, the facet where $[\mathcal{C}]$ shrinks marks a flop transition, and one computes the transformed geometrical data $\kappa'_{abc}$, $c'_a$ and GV invariants of a new Calabi-Yau $X'$ using \eqref{eq:flop_formulas_kappa_c2'} and \eqref{eq:flop_formulas_GV}, and adjoins a new geometrical chamber equal to the dual of $\mathcal{M}_{X'}^{\text{GV}}$.
	\item If, on the other hand, there exists a $\vec{d}$ such that $\kappa_{abc}d^a t^b t^c\equiv 0$ along the dual facet of $\mathcal{K}_X$, one computes the map $w$ of \eqref{eq:Weyl-reflection}, and tests whether $\kappa_{mnl}{w^m}_a {w^n}_b {w^l}_c$ is equal to $\kappa_{abc}'$ computed from \eqref{eq:flop_formulas_kappa_c2'}. If so, one has proven that the newly encountered geometrical chamber is gauge-equivalent to the original one, either via a Weyl reflection or via a flop transition between two diffeomorphic Calabi-Yau's, and can thus be omitted. Otherwise, one records the new phase $X'$, and adjoins a new geometrical chamber equal to the dual of $\mathcal{M}_{X'}^{\text{GV}}$.
\end{enumerate}
For every geometric chamber thus obtained one reruns the above algorithm, and further omits geometric chambers that have previously been generated. Finally, one quotients the result by whatever subgroup of the symmetry group of the birational equivalence class $[X]_\text{b}$ one has access to.

Three kinds of gauge-redundancies remain in $\mathcal{K}_{\circ}$:
\begin{itemize}
	\item $\mathcal{N}=4$ Weyl reflections.
	\item neighboring phases $X$ and $X'$ may still be diffeomorphic if the basis change of $H^2(X,\mathbb{Z})$ needed to make the equivalence of triple intersection numbers and second Chern class manifest, is not equal to a simple reflection across their joined facet of K\"ahler cones.
	\item non-neighboring $X$ and $X'$ may be diffeomorphic, but not related to each other by a known symmetry of $[X]_\text{b}$.
\end{itemize}

\subsection{Limitations of a finite cutoff}  \label{sec:appendixfinite}
In applying a computation of the GV invariants up to a finite degree $d$ to characterize a ray $r$, several misidentifications are possible.
For each possible error we denote by $\mathcal{T}_{\mathrm{hyp}}^{(0)}$ the resulting approximation to the true $\mathcal{T}_{\mathrm{hyp}}$.
The relevant possibilities are:

\begin{enumerate}
  \item A ray $r$ is effective and in $\mathcal{M}_{\infty}$ but is misidentified as being non-effective, because the first nonzero GV invariant along $r$ has degree $>d$. In this case $\mathcal{T}_{\mathrm{hyp}}^{(0)} \supsetneq \mathcal{T}_{\mathrm{hyp}}$.
  \item A ray $r$ is potent but is misidentified as being nilpotent and containing a nop curve, because the infinite series of nonzero GV invariants begins at degree $>d$. In this case $\mathcal{T}_{\mathrm{hyp}}^{(0)} \supsetneq \mathcal{T}_{\mathrm{hyp}}$.
  \item A ray $r$ is nilpotent and on the boundary of $\mathcal{M}_{\infty}$ but is misidentified as containing a nop curve. In this case $\mathcal{T}_{\mathrm{hyp}}^{(0)} \supsetneq \mathcal{T}_{\mathrm{hyp}}$.
  \item A ray $r$ is nilpotent and contains a nop curve but is misidentified as being potent, because the GV invariants are nonzero up to degree $d$.
      In this case $\mathcal{T}_{\mathrm{hyp}}^{(0)} \subsetneq \mathcal{T}_{\mathrm{hyp}}$.
  \item A ray $r$ is nilpotent and contains a nop curve but is misidentified as being on the boundary of $\mathcal{M}_{\infty}$. In this case $\mathcal{T}_{\mathrm{hyp}}^{(0)} \subsetneq \mathcal{T}_{\mathrm{hyp}}$.

\end{enumerate}

In practice, by repeating our computation with ever-increasing $d$, we have reduced the incidence of each of these errors.

\subsection{Checking the lattice WGC up to a finite cutoff}

Given the set $\mathfrak{C}$ of charges with nonzero GV invariants, the set $\mathfrak{F}$ of nilpotent curves outside of $\mathcal{M}_{\infty}$, and all of their associated GV invariants, we can reconstruct an approximation $\mathcal{K}_{\circ}^{(0)}$ to the cone  $\mathcal{K}_{\circ}$ using the methods described in \S\ref{sec:nops} and \S\ref{sec:symmetries}.  We then construct
$\mathcal{T}_{\circ}^{(0)} := \mathscr{T}(\mathcal{K}_{\circ}^{(0)})$.
To check the lattice WGC, we compute points in $\text{Conv}\bigl(\mathcal{T}_{\circ}^{(0)}\bigr)$ subject to the condition
\begin{align}
\vec{p} \cdot \vec{v}_g \leq d_{\mathrm{cutoff}}\,.
\end{align}
Here, as defined above, $\vec{v}_g$ is the grading vector and $d_{\mathrm{cutoff}}$ is the chosen cutoff degree.
We then compute the GV invariant for each curve class $\vec{p}$. If all such GV invariants are nonzero, we conclude that the lattice WGC is satisfied
in $\mathcal{T}_{\circ}^{(0)}$ up to the cutoff $d_{\mathrm{cutoff}}$.
Because $\mathcal{T}_{\circ}^{(0)}$ is a fundamental domain for $\mathcal{T}_{\text{hyp}}$ under (a subgroup of) the automorphism group, we can thereby conclude that the lattice WGC is satisfied in all of $\mathcal{T}_{\text{hyp}}$ up to the specified cutoff.\footnote{Note that if $\mathcal{T}_{\circ}^{(0)}$ is not convex then we may have actually checked a region that is somewhat larger than $\mathcal{T}_{\text{hyp}}$, but not necessarily as large as $\text{Conv}\bigl(\mathcal{T}_{\text{hyp}}\bigr)$,
since the union of the convex hulls of a set of regions is sometimes smaller than the convex hull of the union of the regions.}

\section{Toric Technology and Notation}\label{app:ToricTech}

In this appendix, we review some of the necessary technology that underlies both our general scans as well as the examples discussed in \S\ref{sec:example} and Appendix~\ref{app:examples}, simultaneously setting notation for these examples.

Following Batyrev \cite{Batyrev:1994pg}, we describe Calabi-Yau hypersurfaces in simplicial toric fourfolds, constructed in the following standard way from the four-dimensional reflexive polytopes classified by Kreuzer and Skarke \cite{Kreuzer:2000xy}. The cones over the faces of a reflexive polytope $\Delta^\circ\subset \mathbb{Z}^4$ define a toric fan of a singular toric variety. A fine, regular and star triangulation (FRST) $\mathbb{T}$ of $\Delta^\circ$ defines a desingularized toric fourfold $V_{\Delta^\circ,\mathbb{T}}$ with toric fan assembled from the cones over the simplices of $\mathbb{T}$. The generic anticanonical hypersurface $X_{\Delta^\circ,\mathbb{T}}$ is smooth and Calabi-Yau. The toric divisors associated with points strictly interior to facets do not intersect $X_{\Delta^\circ,\mathbb{T}}$, and thus we will in practice be concerned with triangulations that ignore such points.
Thus, each integer point $p\in \Delta^\circ$ not interior to a facet and other than the origin defines a generator of the Cox ring $x_p$ of homogeneous coordinates, and we may index the generators as $\{x_I\}_{I=1}^{h^{1,1}+4}$ with $h^{1,1}\equiv h^{1,1}(V_{\Delta^\circ,\mathbb{T}})$, and likewise for the points $p_I$. The linear relations among the $p_I$ define the toric $\mathbb{C}^*$-scaling weights of the homogeneous coordinates $x_I$, conveniently organized in a GLSM charge matrix ${Q^a}_I$ whose rows are a $\mathbb{Z}$-basis of linear relations,
\begin{equation}
	\sum_I {Q^a}_I p_I=0\, , \quad a=1,\ldots,h^{1,1}\, .
\end{equation}
We have
\begin{equation}\label{eq:toric_scaling_rel_continuous}
	[x_1:\ldots:x_{h^{1,1}+4}]=\left[\prod_{a=1}^{h^{1,1}}\lambda_a^{{Q^a}_1}x_1:\ldots:\prod_{a=1}^{h^{1,1}}\lambda_a^{{Q^a}_{h^{1,1}+4}}x_{h^{1,1}+4}\right]\, ,\quad \forall \lambda\in (\mathbb{C}^*)^{h^{1,1}}\, ,
\end{equation}
parameterizing the part of the group of toric scaling relations $G_{\text{toric}}$ that is continuously connected to the identity. The full $G_{\text{toric}}$ consists of $\eta^I$ such that $\sum_I \eta^I p_I\in \mathbb{Z}^4$, leading to equivalence relations
\begin{equation}\label{eq:toric_scaling_rel}
	[x_1:\ldots:x_{h^{1,1}+4}]=[e^{2\pi i \eta^1}x_1:\ldots:e^{2\pi i \eta^{h^{1,1}+4}}x_{h^{1,1}+4}]\, .
\end{equation}
We will denote the toric divisors of the toric fourfold by $\hat{D}_I:=\{x_I=0\}\subset V_{\Delta^\circ,\mathbb{T}}$, and the prime toric divisors of the Calabi-Yau threefold $D_I:=\hat{D}_I\cap X_{\Delta^\circ,\mathbb{T}}$. The $\hat{D}_I$ generate the divisor lattice $H^2(V_{\Delta^\circ,\mathbb{T}},\mathbb{Z})$, as well as the cone of effective divisors $\mathcal{E}(V_{\Delta^\circ,\mathbb{T}})$. A choice of ${Q^a}_I$ is equivalent to a choice of basis of curve classes in $H_2(V_{\Delta^\circ,\mathbb{T}},\mathbb{Z})$ whose intersection pairing with the toric divisors $\hat{D}_I$ are the rows of the GLSM charge matrix, and we will specify a basis of curve classes in this manner. We will fix the basis of divisor classes to be its dual. Cones in $\mathbb{T}$ are in one-to-one correspondence with torus-invariant subvarieties equal to the simultaneous vanishing of the homogeneous coordinates associated with one-dimensional sub-cones. All subsets of homogeneous coordinates not associated with a cone in $\mathbb{T}$ in this way generate the Stanley-Reisner (SR) ideal.

All our examples will be favorable in the sense that the natural inclusion $H_2(X_{\Delta^\circ,\mathbb{T}},\mathbb{Z})\hookrightarrow H_2(V_{\Delta^\circ,\mathbb{T}},\mathbb{Z})$ is an isomorphism. The basis of curve and divisor classes in $X_{\Delta^\circ,\mathbb{T}}$ will be the one induced from this isomorphism. We note that different FRSTs of $\Delta^\circ$ lead to Calabi-Yau hypersurfaces in the same birational equivalence class. If a pair of FRSTs have equivalent induced triangulations on all two-faces of $\Delta^\circ$, the corresponding Calabi-Yau hypersurface remains smooth as one transitions from one FRST to the other. We will make use of $\mathcal{K}^\cup_X$ defined as the union over all K\"ahler cones of ambient varieties $V_{\Delta^\circ,\mathbb{T}}$ obtained from the same polytope $\Delta^\circ$, but from distinct FRSTs $\mathbb{T}$ that agree on two-faces. We have $\mathcal{K}^\cup_X\subseteq \mathcal{K}_X$. Furthermore, we define $\mathcal{M}^\cap_X\supseteq \mathcal{M}_X$ as its dual cone.

Finally, we will often drop the subscripts $(\Delta^\circ,\mathbb{T})$ and refer to the toric variety simply as $V$ and the Calabi-Yau hypersurface as $X$, when no confusion is likely to arise.

\section{Examples}\label{app:examples}

In this appendix we work out a few examples in detail that illuminate features we have discussed abstractly in \S\ref{sec:modulispace}. Some of the required toric technology with our chosen notational conventions can be found in Appendix~\ref{app:ToricTech}.

\subsection{Additional constraints on $\mathscr{C}_{\text{BH}}$}  \label{sec:CBHcons}

First, we discuss some properties of the cone of BPS black holes $\mathscr{C}_{\mathrm{BH}}$ that will be relevant in the following examples. Although we placed several inner bounds on it, this cone is generally hard to compute exactly for two reasons. Firstly, it may be generated by non-spherically-symmetric solutions, which are difficult to construct explicitly. Secondly, even upon restricting our attention to the cone of \emph{spherically symmetric} BPS black holes $\mathscr{C}_{\text{BH}}^{\text{spherical}} \subseteq \mathscr{C}_{\mathrm{BH}}$, the resulting attractor flows can run into strongly-coupled CFT boundaries (see \cite{Alim:2021vhs} for more information), which makes it difficult to determine whether a smooth horizon exists or not.

In our main analysis, we have sidestepped these issues by only relying on the inclusion $\mathcal{T}_{\text{hyp}} \subseteq \mathscr{C}_{\mathrm{BH}}$ in \eqref{eq:ThypCBH} when checking the lattice WGC. Here we will be more specific to better understand the following examples.

At any given point in the moduli space $T_a^\star$, the cone of BPS black holes $\mathcal{C}_{\text{BH}}(T_a^\star)$ certainly includes the spherically symmetric solutions that can be explicitly constructed using BPS attractor flows that avoid strongly-coupled CFT boundaries. This region is precisely
\begin{equation}
\mathcal{C}_{\text{BH}}^{\text{explicit}}(T_a^\star) = \Vis(T_a^\star) \,,
\end{equation}
where $\Vis(T_a^\star)$ denotes the region of the Weyl-extended dual-coordinate moduli space that is ``visible'' from $T_a^\star$, i.e., that can be reached from $T_a^\star$ along a straight line in dual coordinates without crossing any boundaries of the Weyl-extended moduli space. Note that $\Vis(T_a^\star)$ is not necessarily contained in $\mathcal{T}_{\text{hyp}}$, i.e., we can freely cross through unstable Weyl flops; the low-energy effective field theory remains under control at these flops despite the wall-crossing phenomena that occur in the ultraviolet BPS spectrum.

Thus, since $\mathcal{C}_{\text{BH}}^{\text{explicit}}(T_a^\star) \subseteq \mathcal{C}_{\text{BH}}(T_a^\star)$, we can strengthen the inclusion \eqref{eq:ThypCBH} to
\begin{equation}
\Vis(\mathcal{T}_{\text{hyp}}) \subseteq \mathscr{C}_{\text{BH}} \,,
\end{equation}
where $\Vis(\mathcal{T}_{\text{hyp}}) \df \bigcup_{T_a^\star \in \mathcal{T}_{\text{hyp}}} \Vis(T_a^\star)$.

Note that, as the fully Weyl-extended dual-coordinate moduli space is not in general convex, $\mathcal{C}_{\text{BH}}^{\text{explicit}}(T_a^\star) = \Vis(T_a^\star)$ generically \emph{does} depends on $T_a^\star$. This does not contradict the conjecture~\eqref{comeonconjecture}, because the inclusion $\mathcal{C}_{\text{BH}}^{\text{explicit}}(T_a^\star) \subseteq \mathcal{C}_{\text{BH}}(T_a^\star)$ need not be strict. If there are no CFT boundaries at codimension one in the dual-coordinate moduli space then we can say a bit more. In this case, $\mathcal{C}_{\text{BH}}^{\text{spherical}}(T_a^\star)$ and $\mathcal{C}_{\text{BH}}^{\text{explicit}}(T_a^\star)$ are essentially the same---at most up to a measure-zero set of points. If the fully Weyl-extended dual-coordinate moduli space still fails to be convex, as occurs in the example discussed in~\S\ref{app:branch_cuts}, then for some $T_a^{\star}$,
\begin{equation}
\mathcal{C}_{\text{BH}}^{\text{spherical}}(T_a^\star) \subsetneq \mathscr{C}_{\text{BH}}^{\text{spherical}} \df \bigcup_{T_a^\star \in \mathcal{T}_{\text{hyp}}} \mathcal{C}_{\text{BH}}^{\text{spherical}}(T_a^\star) \,.
\end{equation}
Given the existence of such examples, we conclude that in general $\mathcal{C}_{\text{BH}}^{\text{spherical}}(T_a^\star) \ne \mathscr{C}_{\text{BH}}^{\text{spherical}}$. This once again does not contradict~\eqref{comeonconjecture} because the cone of BPS black holes may be generated by solutions that are not spherically symmetric. However, settling this question would require constructing all such solutions and  characterizing their properties, a highly non-trivial problem that we leave to future work.

\subsection{An unstable Weyl flop with fully-determined $\mathscr{C}_{\text{BH}}$}\label{app:exact}

We begin with a simple example that contains an unstable Weyl flop, but no CFT boundaries. As a consequence, there are no indeterminate flows and we can determine $\mathscr{C}_{\text{BH}}^{\text{spherical}}$ exactly: it turns out to be a strict subcone of the infinity cone, $\mathcal{M}_{\infty}$. Moreover, while we lack the tools to check this explicitly, the general features of the computed GV invariants suggest that $\mathscr{C}_{\text{BH}} = \mathscr{C}_{\text{BH}}^{\text{spherical}}$ in this example.

The geometry consists of a Calabi-Yau threefold hypersurface with $h^{1,1}=2$ in a toric variety, constructed from an FRST of the reflexive polytope $\Delta^\circ$ whose points other than the origin are
\begin{equation}
	\begin{pmatrix}
		-1&  0&  0&  0&  1&  2\\
		-1&  0&  0&  1&  0&  0\\
		 0&  0&  1&  0&  0& -1\\
		 0&  1&  0&  0&  0& -1
	\end{pmatrix}\, .
\end{equation}
A GLSM charge matrix is given by
\begin{equation}
	\begin{bmatrix}
		 x_1 & x_2 & x_3 & x_4 & x_5 & x_6\\ \hline
		 0 &  1 &  1 &  0 & -2 &  1\\
		 1&  0&  0 &  1 &  1 &  0
	\end{bmatrix} \,.
\end{equation}
The faces of $\Delta^\circ$ are simplices, so no FRST needs to be specified.

In the GLSM basis,  the Mori cone $\mathcal{M}_V$ inherited from the toric ambient variety is simply the first quadrant. The GV invariants of both generators are nonzero, and thus $\mathcal{M}_X=\mathcal{M}_V$ and the K\"ahler cone is given by $\mathcal{K}_X = \{ t^1 \,, t^2  \geq 0 \}$.  The geometry has no flop transitions, so $\mathcal{K} = \mathcal{K}_X$.

The geometry in question has independent triple intersection numbers
\begin{equation}
	\kappa_{111} =  0\,,~~~\kappa_{112} = 3\,,~~~\kappa_{122}=7\,,~~~\kappa_{222} = 14 \,.
\end{equation}
These triple intersection numbers lead to the five-dimensional prepotential
\begin{equation}
	\mathcal{F} = \frac{3}{2} (t^1)^2 t^2 + \frac{7}{2} t^1 (t^2)^2 + \frac{7}{3} (t^2)^3\,.
\end{equation}
The boundary $t^2 \rightarrow 0$ is an asymptotic boundary, which lies at infinite distance. The boundary $t^1 = 0$ is the locus of a genus zero Weyl reflection, where the toric divisor $D_6$ shrinks to a $\mathbb{P}^1$. The Weyl reflection acts on the K\"ahler coordinates as
\begin{equation}
	w:~~ t^1 \rightarrow -t^1\,,~~~ t^2 \rightarrow t^2 + t^1 \,.
	\label{wex3}
\end{equation}
Note that this Weyl reflection is unstable, and correspondingly there are no nop curves in this geometry.  Accordingly, the hyperextended K\"ahler cone is equal to the extended K\"ahler cone, $\mathcal{K}_{\text{hyp}} = \mathcal{K} = \mathcal{K}_X$.

\begin{table}
	\scriptsize \renewcommand{\arraystretch}{0.9}
	\begin{align*}
		\begin{array}{c|ccccccccc}
			\mathdiagbox[width=0.7cm,height=0.5cm,innerleftsep=0.1cm,innerrightsep=0cm]{q_1}{q_2} & 0 & 1 & 2 & 3 & 4 & 5 & 6  & 7 & 8  \\ \hline
			0 & \cellcolor{pink}  *&   \cellcolor{pink}  177&  \cellcolor{pink}  177& \cellcolor{pink}    186&  \cellcolor{pink}   177& \cellcolor{pink}  177 &  \cellcolor{pink}  186&  \cellcolor{pink}   177 & \cellcolor{pink}  177 \\
			1 & \cellcolor{yellow}  	-2	& \cellcolor{green} 178	 &\cellcolor{pink}  20291&\cellcolor{pink} 	317172&	\cellcolor{pink} 2998628&\cellcolor{pink} 	21195310 &\cellcolor{pink} 	123413576&\cellcolor{pink} 	622393836	&\cellcolor{pink} 2806637500 \\
			2&\cellcolor{yellow}  0	&\cellcolor{yellow}  3 &\cellcolor{green} 	-177 &\cellcolor{green}  	332040	&\cellcolor{pink}  73458379 &\cellcolor{pink} 	3048964748 &\cellcolor{pink}  67638465983 &\cellcolor{pink}  1034258133329 &\cellcolor{pink}  12232084778113  \\
			3&\cellcolor{yellow}   0	 &\cellcolor{yellow}  5 &\cellcolor{yellow}  -708 &\cellcolor{green}  44790 &\cellcolor{green}  794368 &\cellcolor{green}  3122149716 &\cellcolor{pink}  710345698242	&\cellcolor{pink}  46445530268176 &\cellcolor{pink}  1663087069097865 \\
			4&\cellcolor{yellow}   0	 &\cellcolor{yellow}  7 &\cellcolor{yellow}  -1068 &\cellcolor{yellow}  75225 &\cellcolor{green}  -4468169 &\cellcolor{green}  243105088 &\cellcolor{green}  54329854510 &\cellcolor{green}  46884487081241 &\cellcolor{pink}  10524250865224651\\
			5&\cellcolor{yellow}   0	 &\cellcolor{yellow}  9 &\cellcolor{yellow}  -1448 &\cellcolor{yellow}  110271 &\cellcolor{yellow}  -7157586 &\cellcolor{green}  396368217 &\cellcolor{green}  -27580928924 &\cellcolor{green}  2382035587157 &\cellcolor{green}  1540781601550297 \\
			6&\cellcolor{yellow}   0	 &\cellcolor{yellow}  11 &\cellcolor{yellow}  -1880 &\cellcolor{yellow}  157734 &\cellcolor{yellow}  -11253268 &\cellcolor{yellow}  676476353 &\cellcolor{green}  -48092153649	 &\cellcolor{green}  2530899579921 &\cellcolor{green} 	-241894701950815 \\
			 7 &\cellcolor{yellow}  0	&\cellcolor{yellow}  13 &\cellcolor{yellow}  	-2412	&\cellcolor{yellow}  231979	&\cellcolor{yellow}  -18701330	&\cellcolor{yellow}  1241479305	&\cellcolor{yellow}  -87415077360	&\cellcolor{green}  4793679740747 &\cellcolor{green}  	-439028227820944 \\
			8 &\cellcolor{yellow}   0	&\cellcolor{yellow}  15	 &\cellcolor{yellow}  -3122	 &\cellcolor{yellow}  356005 &\cellcolor{yellow}  	-32878062 &\cellcolor{yellow} 	2432078638	&\cellcolor{yellow}  -172868371620	&\cellcolor{yellow}  10041154974639	&\cellcolor{green}  -797065258455869
		\end{array}
	\end{align*}
\caption{Genus zero GV invariants $n_{q_1, q_2}^0$ for the geometry of \S\ref{app:exact}. Charges inside the infinity cone $\mathcal{M}_\infty = \left\{ q_1, q_2 \geq 0 \right\}$ but outside the cone of spherically-symmetric BPS black holes $\mathscr{C}_{\text{BH}}^{\text{spherical}} = \left\{  q_2 \geq q_1 \geq 0 \right\}$ are marked in yellow, charges inside $\mathscr{C}_{\text{BH}}^{\text{spherical}}$ but outside the cone of dual coordinates $\mathcal{T} = \left\{  q_2 \geq 2 q_1 \geq 0 \right\}$ are shown in green, and charges inside $\mathcal{T}$ are shown in pink. The absence of zeros within $\mathscr{C}_{\text{BH}}^{\text{spherical}}$ is consistent with the lattice WGC.}
\label{GVextab3}
\end{table}

The GV invariants for this geometry are shown in Table \ref{GVextab3}. The infinity cone $\mathcal{M}_\infty$ is simply equal to the Mori cone, consistent with the relation $\mathcal{M}_\infty = \mathcal{K}_{\text{hyp}}^\vee$.

The dual coordinates $T_a$ are given by
\begin{equation}
	T_1 = 3  t^1 t^2 + \frac{7}{2}  (t^2)^2  \,,~~~T_2 = \frac{3}{2} (t^1)^2 + 7 t^1 t^2 + 7 (t^2)^2 \,.
\end{equation}
These parametrize the cone of dual coordinates,
\begin{equation}
\mathcal{T} = \left\{ T_2 \geq 2 T_1 \geq 0  \right \}\, ,
\end{equation}
which is indeed dual to the effective cone, generated by the divisor classes $[D_5]$ and $[D_6]$.

Under the Weyl reflection, these coordinates transform as $T_1 \rightarrow  T_2 - T_1$, $T_2 \rightarrow T_2$, and the cone of dual coordinates transforms to
\begin{equation}
w(\mathcal{T}) = \left\{  2 T_1 \geq T_2 \,,~~~  T_2 \geq T_1 \right \}\,.
\end{equation}
Since the Weyl-extended dual-coordinate moduli space $\mathcal{W}(\mathcal{T}) = \mathcal{T} \cup w(\mathcal{T})$ is (trivially) convex, the cone of spherically-symmetric BPS black holes is simply
\begin{equation}
\mathscr{C}_{\text{BH}}^{\text{spherical}} = \Vis(\mathcal{T}_{\text{hyp}}) = \mathcal{W}(\mathcal{T}) = \{ T_2 \geq T_1 \geq 0  \} \,.
\end{equation}
As can be seen in Table~\ref{GVextab3}, all the GV invariants within this cone are nonzero up to the degree calculated, in agreement with the lattice WGC.

Finally, we note that the genus zero GV invariants in Table~\ref{GVextab3} have different characteristics inside and outside $\mathscr{C}_{\text{BH}}^{\text{spherical}}$. For example,    their sign strictly alternates by column outside $\mathscr{C}_{\text{BH}}^{\text{spherical}}$, whereas they are mostly positive within $\mathscr{C}_{\text{BH}}^{\text{spherical}}$.
These and other patterns suggest that different physics contributes inside and outside $\mathscr{C}_{\text{BH}}^{\text{spherical}}$, which is perhaps indicative that
$\mathscr{C}_{\text{BH}}^{\text{spherical}}$ may be the full cone of BPS black holes $\mathscr{C}_{\text{BH}}$ in this example.

\subsection{$SU(2)$ enhancement with fundamental matter}\label{app:fundamentally_charged}

Our next example illustrates two new phenomena: (1) the $\mathfrak{su}(2)$ gauge theory arising at a Weyl flop can have massless, fundamentally charged matter and (2) even though it is typically finitely generated, the Weyl group need not be a finite group because different Weyl flops need not commute (even in the absence of a higher-rank nonabelian enhancement).

We consider the toric variety $V$ and its Calabi-Yau hypersurface $X$ arising from the four-dimensional reflexive polytope $\Delta^\circ$ whose points not interior to facets and other than the origin are the columns of
\begin{equation}
\begin{pmatrix}
1& -3&  0& -1&  0& -2&  0\\
0&  1&  0& -1&  0& -1&  1\\
0& -2&  0&  1&  1&  0&  0\\
0& -1&  1&  0&  0&  0&  0
\end{pmatrix}\, .
\end{equation}
A GLSM charge matrix is given by
\begin{equation}\label{eq:GLSM}
\begin{bmatrix}
x_1 & x_2 & x_3 & x_4 & x_5 & x_6 & x_7\\ \hline
1 & 0 & 0 & 1 & -1 & 0 & 1\\
1 & 0 & 0 & -1&  1 & 1 & 0\\
3 & 1 & 1 & 0 &  2 & 0 & -1
\end{bmatrix}\, .
\end{equation}
Evidently, we have $h^{1,1}=3$. All FRSTs of $\Delta^\circ$ give are equivalent along two-faces. We choose, arbitrarily, one such FRST, whose SR ideal is generated by the monomials
\begin{equation}
SR=\langle x_4x_7, x_1x_5x_6, x_1x_6x_7, x_2x_3x_4, x_2x_3x_5 \rangle\, .
\end{equation}
The independent triple intersection numbers $\kappa_{abc}$ are
\begin{align}
&\kappa_{111}=\kappa_{112}=\kappa_{113}=\kappa_{123}=\kappa_{133}=\kappa_{222}=\kappa_{233}=1\nonumber\\
&\kappa_{122}=\kappa_{223}=-1\, ,\quad \kappa_{333}=0\, .
\end{align}
The GV invariants of the generators of $\mathcal{M}_X^\cap$ are all non-vanishing and therefore $\mathcal{M}_X=\mathcal{M}_X^\cap$. The generators, denoted $\{[\mathcal{C}^i]\}_{i=1}^4\subset H_2(X,\mathbb{Z})$, expressed in the basis \eqref{eq:GLSM}, are
\begin{equation}
\begin{pmatrix}
[\mathcal{C}^1]&[\mathcal{C}^2]&[\mathcal{C}^3]&[\mathcal{C}^4]
\end{pmatrix}=
\begin{pmatrix}
1 & 0 & 0 & 1 \\
0 & 1 & 0 & -1\\
0 & 0 & 1 & 1
\end{pmatrix}\, .
\end{equation}
In other words, the K\"ahler cone is defined by $t^a>0$, $a=1,2,3$, and $t^1-t^2+t^3>0$. The two-faces of the Mori cone are generated by the pairs $([\mathcal{C}^i],[\mathcal{C}^j])$ with $(i,j)\in \{(1,2),(2,3),(3,4),(4,1)\}$.

For suitable choice of defining polynomial and K\"ahler class, the Calabi-Yau $X$ has a $\mathbb{Z}_2$ symmetry acting as
\begin{equation}\label{eq:Z2}
s_2:\, H_2(X,\mathbb{Z})\rightarrow H_2(X,\mathbb{Z}):\quad [\mathcal{C}]\mapsto \Lambda\cdot [\mathcal{C}]\, ,\quad \Lambda:=\begin{pmatrix}
0 & 1 & 1\\
1 & 0 & -1\\
0 & 0 & 1
\end{pmatrix}\, ,
\end{equation}
which interchanges $[\mathcal{C}^1] \leftrightarrow [\mathcal{C}^2]$ and $[\mathcal{C}^3] \leftrightarrow [\mathcal{C}^4]$.

In the limit $t^1\rightarrow 0$ the curve $\mathcal{C}_1$ shrinks to a point, and the prime toric divisor $D_5$ shrinks linearly in $t^1$, i.e.~$D_5$ degenerates to a curve. Indeed, $D_5$ is a non-trivial $\mathbb{P}^1$ fibration over $\mathbb{P}^1$ and in the above limit the fiber, in the class $2[\mathcal{C}^1]\in H_2(X,\mathbb{Z})$, shrinks. To see this one notes that the generic anticanonical polynomial $f$ in $V$ can be written, up to overall scale, as
\begin{align}
f=&x_1^2+x_6^2x_7^2\sum_{i=0}^4 g_{8-2i}[x_2:x_3]\cdot (x_4x_5)^i\nonumber\\
&+x_6^3x_7 x_4\sum_{i=0}^3 k_{7-2i}[x_2:x_3]\cdot (x_4x_5)^i
+x_6x_7^3 x_5\sum_{i=0}^3 l_{7-2i}[x_2:x_3]\cdot (x_4x_5)^i\nonumber\\
&+x_6^4 x_4^2\sum_{i=0}^3 m_{6-2i}[x_2:x_3]\cdot (x_4x_5)^i+x_7^4 x_5^2\sum_{i=0}^3 n_{6-2i}[x_2:x_3]\cdot (x_4x_5)^i\, ,
\end{align}
where $\{g_{i},k_i,l_i,m_i,n_i\}$ are generic homogeneous degree $i$ polynomials in $(x_2,x_3)$. Therefore, we have
\begin{equation}\label{eq:P1_fibration}
f|_{x_5=0}= x_1^2+ g_8[x_2:x_3] x_6^2x_7^2 +k_7[x_2:x_3]x_4 x_6^3x_7+m_6[x_2:x_3]x_4^2x_6^2\, .
\end{equation}
Along $D_5$, without loss we may assume $x_6\neq 0$, and can thus gauge fix $x_6=1$.\footnote{Setting $x_6=0$ along $D_5$, the vanishing of $f|_{x_5=x_6}$ would enforce that $x_1=0$. But the monomial $x_1x_5x_6$ is in the Stanley-Reisner ideal of $V$ and thus $x_6\neq 0$ along $D_5$.}
The remaining scaling relations are then summarized in a conveniently reordered charge matrix
\begin{equation}
\begin{bmatrix}
x_2 & x_3 & x_1 & x_4 & x_7\\ \hline
0   &  0  &  1  &  1  &  1 \\
1   &  1  &  3  &  0  &  -1
\end{bmatrix}\, ,
\end{equation}
associated with the three-dimensional toric variety $\hat{D}_5$.

We can therefore view the divisor $D_5$ as a hypersurface in the toric threefold $\hat{D}_5$. One immediately sees that $\hat{D}_5$ is a $\mathbb{P}^2$ fibration over $\mathbb{P}^1$:
\begin{equation}
\mathbb{P}^2_\mathrm{f}\hookrightarrow \hat{D}_5 \twoheadrightarrow \mathbb{P}^1_\mathrm{b}\, ,
\end{equation}
and the polynomial \eqref{eq:P1_fibration} defines a quadratic hypersurface in $\mathbb{P}^2_\mathrm{f}$, i.e. a $\mathbb{P}^1$ in the class $2[H]$ where $[H]$ is the hyperplane class of the fiber, over each point of $\mathbb{P}^1_\mathrm{b}$ parameterized by the projective coordinates $[x_2:x_3]$. Therefore, the divisor $D_5$ is a $\mathbb{P}^1$ fibration over $\mathbb{P}^1$,
\begin{equation}\label{eq:P1-fibration_D5}
\mathbb{P}^2_\mathrm{f} \supset \mathbb{P}^1_{2[H]}\hookrightarrow D_5 \twoheadrightarrow \mathbb{P}^1_\mathrm{b}\, ,
\end{equation}
which degenerates over all points in the base along which the discriminant
\begin{equation}
\Delta_{f}:=4m_6g_8-(k_7)^2
\end{equation}
vanishes. For generic choice of $\{g_8,k_7,m_6\}$ the $\mathbb{P}^1_{2[H]}$ fiber thus degenerates to the union of two $\mathbb{P}^1$'s, each in the hyperplane class $[H]$ of $\mathbb{P}^2_{\mathrm{f}}$, over $14$ points in the base $\mathbb{P}^1$, thus totaling $28$ isolated rigid $\mathbb{P}^1$'s.

The generic $\mathbb{P}^1$ fiber is obtained by intersecting $D_5$ with a generic representative of the class $[D_2]=[D_3]$. By intersecting with a basis of divisors, one computes that it is in the homology class $2[\mathcal{C}^1]\in H_2(X,\mathbb{Z})$. Its curve moduli space $\mathcal{M}_{2[\mathcal{C}^1]}$ is equal to the base of the $\mathbb{P}^1$ fibration \eqref{eq:P1-fibration_D5}, and thus its GV invariant is
\begin{equation}
n^0_{2[\mathcal{C}^1]}=(-1)^{\text{dim}\left(\mathcal{M}_{2[\mathcal{C}^1]}\right)}\chi(\mathcal{M}_{2[\mathcal{C}^1]})=-2\, .
\end{equation}
As in \cite{Aspinwall:1995xy,Katz:1996ht}, M2-branes wrapped on the $\mathbb{P}^1$ fiber become massless W-bosons in the singular limit, and the gauge group enhances an $SU(2)$ factor. The presence of $28$ isolated $\mathbb{P}^1$'s in the class $[\mathcal{C}^1]$ on the other hand implies that
\begin{equation}
n^0_{[\mathcal{C}^1]}=28\, ,
\end{equation}
and in the singular limit M2-branes wrapped on these curves lead to $28$ massless hypermultiplets with half the Cartan-$U(1)$ charge of the W-bosons, which therefore must organize into $14$ hypermultiplets in the fundamental representation of the gauge group $SU(2)$. This direct prediction for the GV invariants is confirmed by an independent systematic computation of GV invariants using mirror symmetry \cite{Hosono:1993qy,Hosono:1994ax,computational-mirror-symmetry}, see Table \ref{GVextab2}.

As usual, their contribution to the prepotential of the type IIA compactification has a logarithmic branch cut starting at the singular locus $Z:=[\mathcal{C}^1]_az^a=z^1=0$,
\begin{equation}
\mathcal{F}^{\text{IIA}}\supset -\frac{1}{(2\pi i)^3} \left(28\,\text{Li}_3(e^{2\pi i Z})-2\,\text{Li}_3(e^{4\pi i Z})\right)\simeq \frac{1}{2}Z^2\left(28-4\times 2\right)\frac{\log(1/Z)}{(2\pi i)}+\text{hol.}\, ,
\end{equation}
leading to running of the $U(1)$ gauge coupling
\begin{equation}
\tau_{U(1)}=\del_Z\left([D_5]^a\del_a \mathcal{F}^{\text{IIA}}\right)= -\frac{i}{2\pi} \times 2b_{\text{YM}}\times \log\left(\frac{\Lambda_\text{L}}{\mathcal{Z}}\right)+\mathcal{O}(\mathcal{Z}/M_P)\, ,
\end{equation}
with
\begin{equation}
b_{\text{YM}}:=\mu\del_\mu \frac{8\pi^2}{g^2_{\text{YM}}(\mu)}=2N_c-N_F=-10\, ,\quad \Lambda_\text{L}:=\mu_{\text{UV}}e^{\frac{2\pi i}{b_{\text{YM}}}\tau_{\text{YM}}(\mu_{\text{UV}})}\, ,\quad \tau_{\text{YM}}(\mu_{\text{UV}})=z^2\, ,
\end{equation}
with UV matching scale $\mu_{\text{UV}}:= M_P e^{\frac{K}{2}}$, where $K$ is defined in \eqref{eq:defkahler} and $\mathcal{Z}:=-2\pi i \mu_{\text{UV}} Z$.
Here, $N_c=2$ is the dual Coxeter number of the gauge group $SU(2)$, $N_F=14$ is the number of adjoint hypermultiplets, $\Lambda_\text{L}$ is the Landau pole of the IR-free gauge theory, and $\ell_5\cdot \tau_{\text{YM}}(\mu_{\text{UV}})$ is the classical gauge coupling of the five-dimensional theory as a function of the remaining K\"ahler parameters.

The Weyl group associated with the $SU(2)$ gauge group is generated by
\begin{equation}
w:\,H_2(X,\mathbb{Z})\rightarrow H_2(X,\mathbb{Z})\, , \quad [\mathcal{C}]\mapsto \begin{pmatrix}
-1 & 2 & 4\\
0 & 1 & 0\\
0 & 0 & 1
\end{pmatrix}\cdot [\mathcal{C}]\, ,
\end{equation}
but the GV invariants turn out to not be invariant under this Weyl group action (cf. Table \ref{GVextab2}), implying a non-trivial wall-crossing phenomenon at the origin of the Coulomb branch.

\begin{figure}
\centering
\begin{tabular}{c c}
\includegraphics[keepaspectratio,width=7cm]{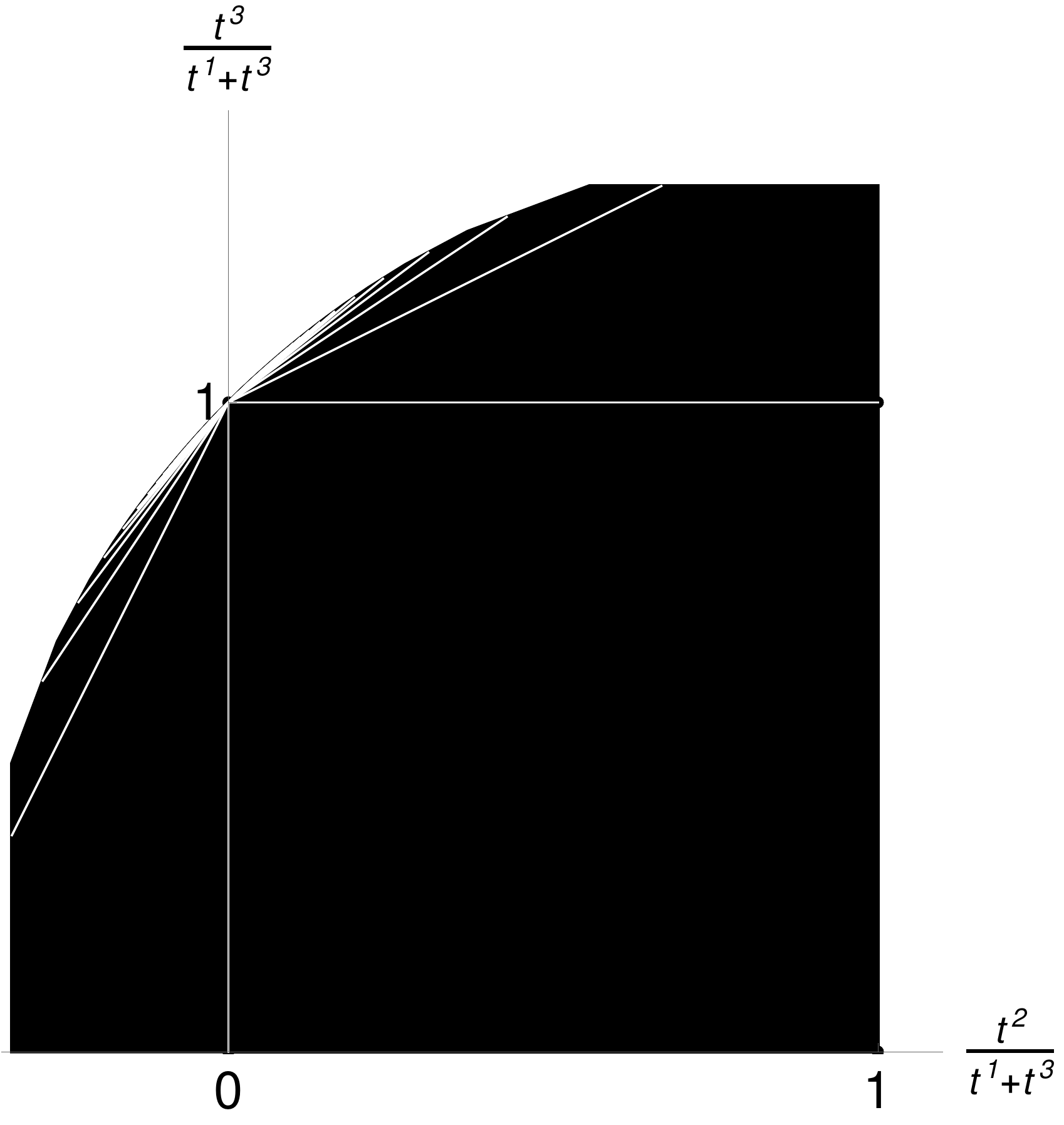} & \includegraphics[keepaspectratio,width=8.4cm]{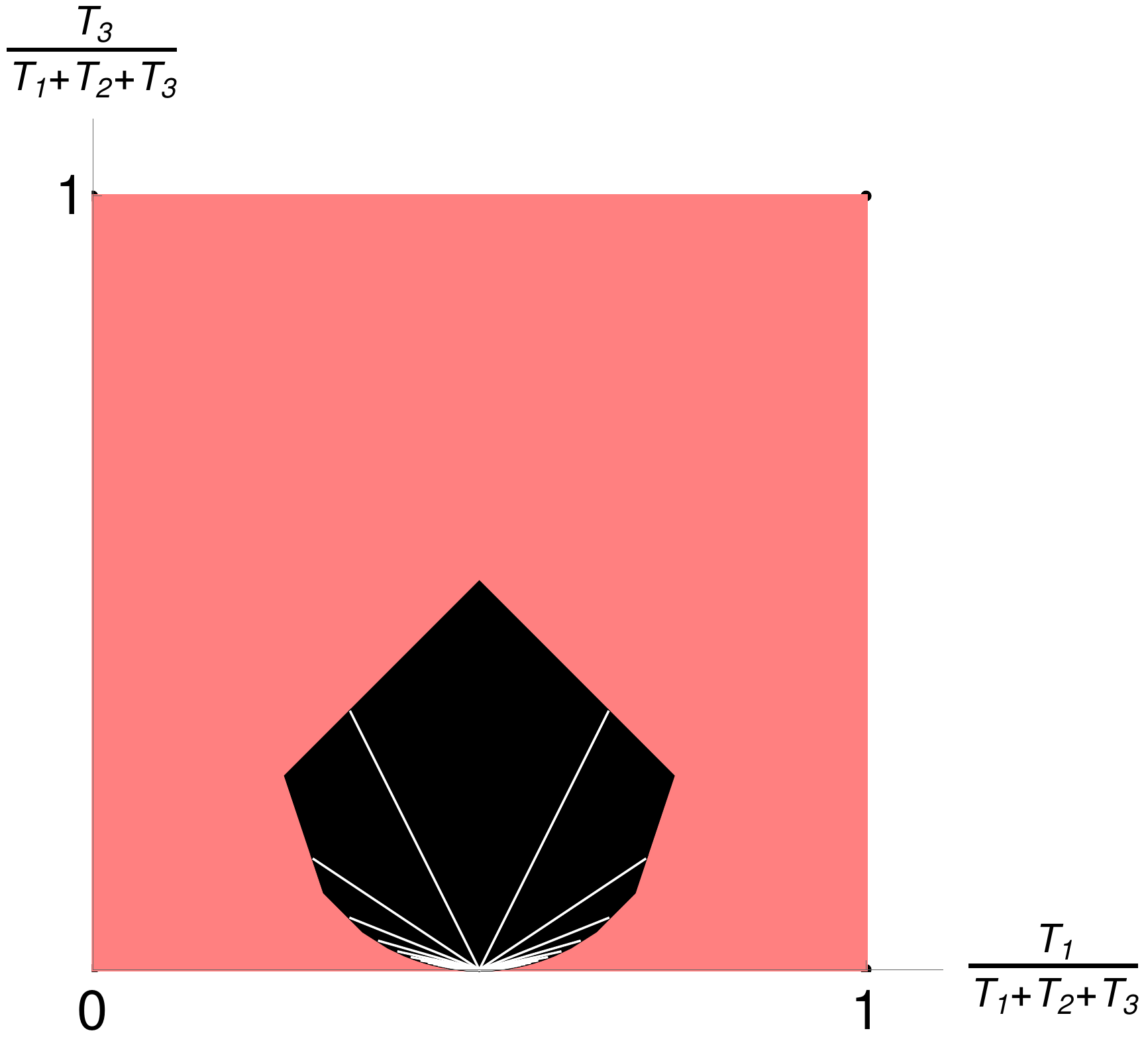}
\end{tabular}
\caption{Left: Two-dimensional cross-section of the Weyl-group orbit of the K\"ahler cone. Interior lines mark Weyl-reflection loci, while exterior walls mark the CFT limits. Right: Two-dimensional cross-section of the Weyl-group orbit of the cone of movable curves (black), embedded into the Mori cone (pink). Again, Weyl-reflection loci are marked by white lines. The lower two vertices are the curves that shrink to give rise to $\mathfrak{su}(2)$ enhancements while the upper two correspond to the CFT limits.}
\label{fig:Weyl}
\end{figure}

The physics described above of course applies equally well to the codimension-one facet of the K\"ahler cone where the curve $\mathcal{C}^2$ shrinks, via the symmetry \eqref{eq:Z2}. As the curve classes $([\mathcal{C}^1],[\mathcal{C}^2])$ span a two-face of the Mori cone, one might expect that the simultaneous shrinking of these two curves leads to higher rank non-abelian enhancement. This, however, is not so: the diagonal limit $(t^1,t^2)\rightarrow 0$ lies at infinite distance in moduli space, where $X$ degenerates into a complex surface. We further note that passing through Weyl reflections associated with the shrinking curves $\mathcal{C}^1$ and $\mathcal{C}^2$ in alternating order, one obtains the infinite-order Weyl group $\mathcal{W}$ generated by
\begin{equation}\label{eq:Weyl_group_generator}
w':=s_2\circ w:\,H_2(X,\mathbb{Z})\rightarrow H_2(X,\mathbb{Z})\, , \quad [\mathcal{C}]\mapsto \begin{pmatrix}
0 & 1 & 1\\
-1 & 2 & 3\\
0 & 0 & 1
\end{pmatrix}\cdot [\mathcal{C}]\, .
\end{equation}
The Weyl orbit of the K\"ahler cone is an infinitely generated convex cone, and the Weyl orbit of the cone of movable curves is likewise an infinitely generated convex sub-cone of the Mori cone (see Figure \ref{fig:Weyl}). All Weyl-group images of the K\"ahler cone intersect at an accumulation point, which is the aforementioned intersection of both non-abelian enhancement loci, lying at infinite distance in the moduli space.

Along the other two facets of the K\"ahler cone the prime toric divisors $D_6$ and $D_7$ shrink to points, so the low energy physics is a strongly coupled tensionless string CFT in the five-dimensional theory.

\begin{table}
	\scriptsize \renewcommand{\arraystretch}{0.9}
	\begin{align*}
		\rotatebox[origin=c]{90}{\underline{$q_3=0$}}
		\begin{array}{c|ccccccc}
			\mathdiagbox[width=0.7cm,height=0.5cm,innerleftsep=0.1cm,innerrightsep=0cm]{q_1}{q_2} & 0 & 1 & 2 & 3 & 4 & 5 & 6   \\ \hline
			0 & \cellcolor{pink} *& \cellcolor{yellow}   28& \cellcolor{yellow}   -2& \cellcolor{yellow}    0& \cellcolor{yellow}    0& \cellcolor{yellow}    0& \cellcolor{yellow}    0\\
			1& \cellcolor{yellow} 28& \cellcolor{yellow}  \cellcolor{pink}  196& \cellcolor{yellow}   28& \cellcolor{yellow}    0& \cellcolor{yellow}    0& \cellcolor{yellow}    0& \cellcolor{yellow}    0\\
			2& \cellcolor{yellow} -2& \cellcolor{yellow}   28& \cellcolor{yellow}  \cellcolor{pink} 176& \cellcolor{yellow}   28& \cellcolor{yellow}   -2& \cellcolor{yellow}    0& \cellcolor{yellow}    0\\
			3& \cellcolor{yellow}  0& \cellcolor{yellow}    0& \cellcolor{yellow}   28& \cellcolor{yellow}  \cellcolor{pink} 196& \cellcolor{yellow}   28& \cellcolor{yellow}    0& \cellcolor{yellow}    0\\
			4& \cellcolor{yellow}  0& \cellcolor{yellow}    0& \cellcolor{yellow}   -2& \cellcolor{yellow}   28& \cellcolor{yellow}  \cellcolor{pink} 176& \cellcolor{yellow}   28& \cellcolor{yellow}   -2\\
			5& \cellcolor{yellow}  0& \cellcolor{yellow}    0& \cellcolor{yellow}    0& \cellcolor{yellow}    0& \cellcolor{yellow}   28& \cellcolor{yellow}  \cellcolor{pink} 196& \cellcolor{yellow}   28\\
			6& \cellcolor{yellow}  0& \cellcolor{yellow}    0& \cellcolor{yellow}    0& \cellcolor{yellow}    0& \cellcolor{yellow}   -2 & \cellcolor{yellow}   28& \cellcolor{yellow}   \cellcolor{pink} 176
		\end{array}
	\end{align*}
	\begin{align*}
		\rotatebox[origin=c]{90}{\underline{$q_3=1$}}
		\begin{array}{c|ccccccc}
			\mathdiagbox[width=0.7cm,height=0.5cm,innerleftsep=0.1cm,innerrightsep=0cm]{q_1}{q_2}   & -1 & 0 & 1 & 2 & 3 & 4 & 5    \\ \hline
			0 & -  & \cellcolor{yellow}  252 & \cellcolor{yellow}    0& \cellcolor{yellow}    0& \cellcolor{yellow}    0& \cellcolor{yellow}    0& \cellcolor{yellow}    0\\
			1 & \cellcolor{yellow} 252& \cellcolor{yellow}\cellcolor{pink} 26144 & \cellcolor{yellow}\cellcolor{pink} 161240& \cellcolor{yellow}34336& \cellcolor{yellow}-16132& \cellcolor{yellow}24576& \cellcolor{yellow}-32768\\
			2 & \cellcolor{yellow} 0  & \cellcolor{yellow}\cellcolor{pink} 161240& \cellcolor{yellow}\cellcolor{pink}3389568& \cellcolor{yellow}\cellcolor{pink} 11865384& \cellcolor{yellow}\cellcolor{green}3618944& \cellcolor{yellow}\cellcolor{green}-297512& \cellcolor{yellow}688128\\
			3 & \cellcolor{yellow} 0& \cellcolor{yellow}34336& \cellcolor{yellow}\cellcolor{pink} 11865384& \cellcolor{yellow}\cellcolor{pink}144023712& \cellcolor{yellow}\cellcolor{pink} 379678864& \cellcolor{yellow}\cellcolor{green}147415200& \cellcolor{yellow}\cellcolor{green}5082408\\
			4 & \cellcolor{yellow} 0& \cellcolor{yellow}-16132& \cellcolor{yellow}\cellcolor{green}3618944& \cellcolor{yellow}\cellcolor{pink} 379678864& \cellcolor{yellow}\cellcolor{pink}3340910720& \cellcolor{yellow}\cellcolor{pink} 7490806028& \cellcolor{yellow}\cellcolor{green}3376234624\\
			5 & \cellcolor{yellow} 0& \cellcolor{yellow}24576& \cellcolor{yellow}\cellcolor{green}-297512& \cellcolor{yellow}\cellcolor{green}147415200& \cellcolor{yellow}\cellcolor{pink} 7490806028& \cellcolor{yellow}\cellcolor{pink}52954926336& \cellcolor{yellow}\cellcolor{pink} 107008961736\\
			6 & \cellcolor{yellow} 0& \cellcolor{yellow}-32768& \cellcolor{yellow}688128&\cellcolor{green}5082408&\cellcolor{green}3376234624&\cellcolor{pink} 107008961736&\cellcolor{pink}643290939648
		\end{array}
	\end{align*}
	\begin{align*}
	\rotatebox[origin=c]{90}{\underline{$q_3=2$}}
	\begin{array}{c|ccccccc}
		\mathdiagbox[width=0.7cm,height=0.5cm,innerleftsep=0.1cm,innerrightsep=0cm]{q_1}{q_2}   & -2 & -1 & 0 & 1 & 2 & 3 & 4    \\ \hline
		0 &  -&    -& \cellcolor{yellow}-9252& \cellcolor{yellow}    0& \cellcolor{yellow}    0& \cellcolor{yellow}    0& \cellcolor{yellow}    0\\
		1 & -& \cellcolor{yellow}    0& \cellcolor{yellow}143640& \cellcolor{yellow}1005480& \cellcolor{yellow}143640& \cellcolor{yellow}    0& \cellcolor{yellow}    0\\
		2 & \cellcolor{yellow} -9252& \cellcolor{yellow}143640&\cellcolor{pink}107389712&\cellcolor{pink}1982306472&\cellcolor{pink}6785861560&\cellcolor{green}2202638504&\cellcolor{green}-365781232\\
		3 & \cellcolor{yellow} 0& \cellcolor{yellow}1005480&\cellcolor{pink}1982306472&\cellcolor{pink}74208835520&\cellcolor{pink}669904166680&\cellcolor{pink}1684741323120&\cellcolor{green}699283955480\\
		4 & \cellcolor{yellow} 0& \cellcolor{yellow}143640&\cellcolor{pink}6785861560&\cellcolor{pink}669904166680&\cellcolor{pink}13132156970304&\cellcolor{pink}82543157015248&\cellcolor{pink}172178353335592\\
		5 & \cellcolor{yellow} 0& \cellcolor{yellow}    0&\cellcolor{green}2202638504&\cellcolor{pink}1684741323120&\cellcolor{pink}82543157015248&\cellcolor{pink}1115085309377792&\cellcolor{pink}5538585837897800\\
		6 & \cellcolor{yellow} 0& \cellcolor{yellow}    0&\cellcolor{green}-365781232&\cellcolor{green}699283955480&\cellcolor{pink}172178353335592&\cellcolor{pink}5538585837897800&\cellcolor{pink}57882723471476816
	\end{array}
	\end{align*}
	\begin{align*}
	\rotatebox[origin=c]{90}{\underline{$q_3=3$}}
	\begin{array}{c|cccccc}
		\mathdiagbox[width=0.7cm,height=0.5cm,innerleftsep=0.1cm,innerrightsep=0cm]{q_1}{q_2}  &-3 & -2 & -1 & 0 & 1 & 2     \\ \hline
		0 &  -&   -&    -& \cellcolor{yellow}848628& \cellcolor{yellow}    0& \cellcolor{yellow}    0\\
		1&  -&    -& \cellcolor{yellow}    0& \cellcolor{yellow}-18865280& \cellcolor{yellow}-132056960&  \cellcolor{yellow} -18865280\\
		2& -& \cellcolor{yellow}    0& \cellcolor{yellow}    0& \cellcolor{yellow}1000128720& \cellcolor{yellow}19800596480& \cellcolor{yellow}69523670960\\
		3& \cellcolor{yellow} 848628& \cellcolor{yellow}-18865280& \cellcolor{yellow}1000128720& \cellcolor{yellow}\cellcolor{pink}1155156240800& \cellcolor{yellow}\cellcolor{pink}37248393860088& \cellcolor{yellow}\cellcolor{pink}317060747387520\\
		4& \cellcolor{yellow} 0& \cellcolor{yellow}-132056960& \cellcolor{yellow}19800596480& \cellcolor{yellow}\cellcolor{pink}37248393860088& \cellcolor{yellow}\cellcolor{pink}1975319069667328& \cellcolor{yellow}\cellcolor{pink}30784544966451680\\
		5& \cellcolor{yellow} 0& \cellcolor{yellow}-18865280& \cellcolor{yellow}69523670960&\cellcolor{pink}317060747387520&\cellcolor{pink}30784544966451680&\cellcolor{pink}832309647138723328
	\end{array}
	\end{align*}
	\begin{align*}
	\rotatebox[origin=c]{90}{\underline{$q_3=4$}}
	\begin{array}{c|cccccc}
		\mathdiagbox[width=0.7cm,height=0.5cm,innerleftsep=0.1cm,innerrightsep=0cm]{q_1}{q_2} & -4  &-3 & -2 & -1 & 0 & 1      \\ \hline
		0 &  -&    -&    -&     -& \cellcolor{yellow}-114265008& \cellcolor{yellow}    0\\
		1& -&   -&     -& \cellcolor{yellow}    0& \cellcolor{yellow}3226808340& \cellcolor{yellow}22587658380\\
		2&  -&    -& \cellcolor{yellow}    0& \cellcolor{yellow}    0& \cellcolor{yellow}-149516774740& \cellcolor{yellow}-2854273434600\\
		3&  -& \cellcolor{yellow}    0& \cellcolor{yellow}    0& \cellcolor{yellow}5158112400& \cellcolor{yellow}12435686082500& \cellcolor{yellow}427372552969920\\
		4& \cellcolor{yellow}-114265008& \cellcolor{yellow}3226808340& \cellcolor{yellow}-149516774740& \cellcolor{yellow}12435686082500&\cellcolor{pink}18251242470992832&\cellcolor{pink}850602661974296708\\
		5& \cellcolor{yellow}0& \cellcolor{yellow}22587658380& \cellcolor{yellow}-2854273434600& \cellcolor{yellow}427372552969920&\cellcolor{pink}850602661974296708&\cellcolor{pink}58131312791048887904
	\end{array}
	\end{align*}
\caption{Genus zero GV invariants $n_{q_1, q_2, q_3}^0$ for the geometry of Appendix \ref{app:fundamentally_charged}. The sites labeled ``$-$'' lie outside the (non-simplicial) Mori cone. The infinity cone is equal to the Mori cone and is marked in yellow. The cone $\mathcal{T}$ is marked in pink and its images under $w'$ and $w'^{-1}$ of eq. \eqref{eq:Weyl_group_generator} are marked in green (higher order images do not have integer sites in their interior that fall into the displayed window). Together, the pink and green sites denote the largest region $\text{Vis}(\mathcal{T}_{\text{hyp}})$ where BPS black holes are known to exist. These charges are fully populated by non-vanishing GV-invariants, in agreement with the lattice WGC.
}
\label{GVextab2}
\end{table}
Absent any flop transitions, we have $\mathcal{K}\equiv \mathcal{K}_X$ and the effective cone is generated by the (rigid) prime toric divisors $\{D_4,D_5,D_6,D_7\}$. Its dual, the cone of movable curves $\text{Mov}$, turns out to be fully populated by non-vanishing genus zero GV invariants, as predicted by the lattice WGC, and is indeed equal to the cone of dual coordinates $\mathcal{T}$.

The hyperextended K\"ahler cone $\mathcal{K}_{\text{hyp}}$ is also equal to $\mathcal{K}$, as all loci of $\mathfrak{su}(2)$ enhancement are unstable. Therefore, by our discussion in \S\ref{sec:modulispace}, the infinity cone $\mathcal{M}_\infty$ should be equal to the entire Mori cone. This indeed appears to be borne out in this example: the sequence of GV invariants associated with either of the two CFT limits is
\begin{align}
	n_{(0,0,k)}^g = &\left\{252,
	-9252,
	848628,
	-114265008,
	18958064400,
	-3589587111852, \right. \nonumber\\
	&\left.
	744530011302420,
	-165076694998001856,
	38512679141944848024, \right. \nonumber \\
	&\left. -9353163584375938364400,
	2346467355966572489025540,\ldots\right\}\, ,
\end{align}
and does not appear to terminate. Likewise, adding arbitrary multiples of one of the generators that shrinks at an $\mathfrak{su}(2)$-enhancement locus to a suitable curve class gives rise to a sequence of GV invariants that appears infinite, e.g.:
\begin{align}
	n_{(1,2,1)+k\cdot (1,0,0)}=&\left\{34336,
	11865384,
	144023712,
	379678864,
	147415200,
	5082408,
	10208800,\right.\nonumber\\
	&\left.-13565952,
	16957440,
	-20348928,
	23740416,
	-27131904,
	30523392,\right.\nonumber\\
	&\left.-33914880,
	37306368,
	-40697856,
	44089344,-47480832,\ldots\right\}\, .
\end{align}

Finally, despite the presence of walls of marginal stability along the Weyl-reflection loci in moduli space, the entire Weyl-orbit $\mathcal{W}(\mathcal{T})$ of the cone of dual coordinates is fully populated by non-vanishing GV invariants, see Table \ref{GVextab2}. Because this region is convex in the present example, this is precisely $\Vis(\mathcal{T}_{\text{hyp}})$, so the absence of any vanishing GV invariants within this larger cone provides a more stringent test of the lattice WGC.

\subsection{Monodromy in the Weyl-extended moduli space}\label{app:branch_cuts}

Our final example involves yet another new feature: branch cuts and monodromy in the fully Weyl-extended moduli space. This example has two Weyl flops, one stable and one unstable, with a CFT boundary at their codimension-two intersection. Upon passing through the two Weyl flops alternately in sequence, one discovers that the CFT point is a branch point for the fully Weyl-extended moduli space, with a monodromy in the central charges as one moves around this branch point. This monodromy is not visible when one restricts attention to the hyperextended moduli space $\mathcal{T}_{\text{hyp}}$ (avoiding the unstable Weyl flop). Indeed, since $\mathcal{K}_{\text{hyp}} = \mathcal{M}_{\text{hyp}}^\vee$ is convex, the $\mathscr{T}$ map is invertible throughout $\mathcal{K}_{\text{hyp}}$, forbidding such a monodromy.

Let us consider the $h^{1,1}=3$ Calabi-Yau threefold hypersurface constructed from an FRST of the reflexive polytope $\Delta^\circ$ whose points not interior to facets, excluding the origin, are the columns of
\begin{equation}
	\begin{pmatrix}
		-9&  0&  0&  1&  1& -4& -2\\
		-5&  0&  1&  0&  1& -2& -1\\
		-3&  1&  0&  0&  1& -1&  0\\
		-2&  0&  0&  0&  2&  0&  0
	\end{pmatrix}\, .
\end{equation}
A GLSM charge matrix is given by
\begin{align}\label{eq:glsm}
	\begin{bmatrix}
		x_1 & x_2 & x_3 & x_4 & x_5 & x_6 & x_7\\ \hline
		1 & 0 & 0 & 0 & 1 & -2 & 0 \\
		0 & 1 & 0 & 0 & 0 & 1 & -2 \\
		0 & 0 & 1 & 2 & 0 & 0 & 1
	\end{bmatrix} \,.
\end{align}
All FRSTs of $\Delta^\circ$ are equivalent along two-faces, so without loss we can choose a single arbitrary FRST of $\Delta^\circ$. The cone $\mathcal{M}^\cap_X$ is the first octant in the basis \eqref{eq:glsm}.

Our geometry has nonzero independent triple intersection numbers
\begin{equation}
	\kappa_{123} =  1\,,~~~\kappa_{223} = \kappa_{133} = 2\,,~~~\kappa_{233}=4\,,~~~\kappa_{333}=8\,,
	\label{extrip}
\end{equation}
while the rest vanish. These triple intersection numbers lead to the 5d prepotential
\begin{equation}
	\mathcal{F} = t^1 t^2 t^3 + (t^2)^2 t^3 + t^1 (t^3)^2 + 2 t^2 (t^3)^2 + 4 (t^3)^3/3\,.
	\label{prepex}
\end{equation}
In the limit $t^3\rightarrow 0$ the Calabi-Yau volume vanishes, so this limit corresponds to a facet of the K\"ahler cone that lies at infinite distance in moduli space. At $t^1=0$ the prime toric divisor $D_6$ shrinks to a curve of genus one, while as $t^2=0$ the prime toric divisor $D_7$ shrinks to a curve of genus zero.\footnote{One sees this from the data of the polytope as follows: the prime toric divisors $D_{6,7}$ are associated to points interior to one-faces of $\Delta^\circ$. Their $\mathbb{P}^1$ fibers --- the curves that shrink at the respective facets of the K\"ahler cone --- are associated to edges of the triangulation interior to two-faces that end in the respective one-face points. The genera of the curves they degenerate to are equal to the genera of the one-faces, defined as the number of points interior to the dual two-faces of the dual polytope.} Along each of these latter two boundaries, there is an $\mathfrak{su}(2)$ enhancement of the gauge symmetry, and there is a Weyl reflection of the moduli space through the boundary. Thus, the Mori cone is simplicial and its generators coincide with our choice of basis.

The genus one Weyl reflection is given by
\begin{equation}
	w_1:~~ t^1 \rightarrow -t^1\,,~~~ t^2 \rightarrow t^2 + t^1\,,~~~t^3 \rightarrow t^3\,,
	\label{w1}
\end{equation}
under which the prepotential in \eqref{prepex} is invariant.
Relatedly, the genus zero GV invariant $n_{1,0,0}^0$ vanishes due to a cancelation between hypermultiplets and vector multiplets, which become massless at the wall $t^1 =0$.

Meanwhile, under the genus zero Weyl reflection,
\begin{equation}
	w_2:~~ t^1 \rightarrow t^1\,,~ t^2 \rightarrow -t^2 \,,~t^3 \rightarrow t^3+t^2\,,
\end{equation}
the prepotential shifts as $\mathcal{F} \rightarrow \mathcal{F} + (t^2)^3/3$. The corresponding shift $\kappa_{222} \rightarrow \kappa_{222} + 2$ can be read off from the GV invariant $n_{0,1,0}^0 = -2$, which tells us that a charged vector multiplet (and no hypermultiplets) becomes massless at $t^2=0$, as expected for a genus zero Weyl reflection.

This geometry has no flop transitions, so the extended K\"ahler cone $\mathcal{K}$ is given simply by $\mathcal{K}_X$.

The dual coordinates are given by $T_a := \frac{1}{2} \kappa_{abc} t^b t^c$, so by \eqref{extrip} they take the form
\begin{equation}
	T_1 = t^2 t^3 + (t^3)^2\,,~~~T_2 = t^1 t^3 + 2 t^2 t^3 + 2 (t^3)^2\,,~~~T_3 = t^1 t^2 + (t^2)^2 + 2 t^1 t^3 + 4 t^2 t^3 + 4 (t^3)^2\,.
\end{equation}
These parametrize the cone of dual coordinates,
\begin{equation}
	\mathcal{T}  = \left \{ T_1 \geq 0\,,~~~T_2 \geq 2 T_1\,,~~~ T_3 \geq 2 T_2    \right \}\,.
\end{equation}
As expected, $\mathcal{T}$ is contained in the Mori cone, $\mathcal{M}_X$, which is by definition the dual of the K\"ahler cone:
\begin{equation}
	\mathcal{M}_X = \mathcal{K}_X^\vee = \left\{ T_1, T_2, T_3 \geq 0 \right\}\,.
\end{equation}
Meanwhile, the effective cone is
\begin{equation}
	\mathcal{E} = \mathcal{T}^\vee =  \left\{ t^1 + 2 t^2 + 4 t^3 \geq 0\,,~t^2 + 2 t^3\geq 0\,,~t^3 \geq 0 \right\} \,,
\end{equation}
and in this case is generated by the prime toric divisors $\{D_1,D_6,D_7\}$.

\begin{figure}
	\centering
	\includegraphics[keepaspectratio,width=10cm]{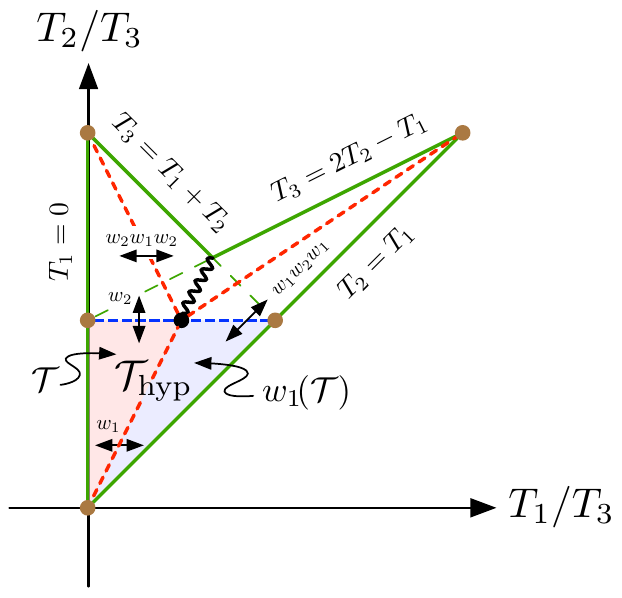}
	\caption{Regions in the space of dual coordinates. The hyperextended cone of dual coordinates $\mathcal{T}_{\rm hyp}$ is the union of the cone of dual coordinates $\mathcal{T}$ (shaded dusty rose) and its image under the genus one Weyl reflection, $w_1(\mathcal{T})$, shaded lavender. Additional regions of moduli space may be accessed by further Weyl reflections, and the (non-convex) part of moduli space $\text{Vis}(\mathcal{T}_{\text{hyp}})$ that may be connected by straight lines through moduli space to $\mathcal{T}_{\rm hyp}$ is bounded in green. By the analysis of \cite{Alim:2021vhs}, BPS black holes necessarily exist in $\text{Vis}(\mathcal{T}_{\text{hyp}})$. }
	\label{fig:vis}
\end{figure}

\begin{table}
\scriptsize \renewcommand{\arraystretch}{0.9}
\begin{align*}
\rotatebox[origin=c]{90}{\underline{$q_1=0$}}
\begin{array}{c|ccccccc}
      \mathdiagbox[width=0.7cm,height=0.5cm,innerleftsep=0.1cm,innerrightsep=0cm]{q_2}{q_3}   & 0 & 1 & 2 & 3 & 4 & 5 & 6   \\ \hline
  0& \cellcolor{pink}       *& \cellcolor{pink}            96& \cellcolor{pink}             144& \cellcolor{pink}              96& \cellcolor{pink}             144& \cellcolor{pink}              96& \cellcolor{pink}             144 \\
   1& \cellcolor{yellow}           -2& \cellcolor{green}              96& \cellcolor{pink}            5384& \cellcolor{pink}           69376& \cellcolor{pink}          631788& \cellcolor{pink}         4247296& \cellcolor{pink}        24219184  \\
  2& \cellcolor{yellow}                           0& \cellcolor{yellow}               0& \cellcolor{green}             144& \cellcolor{green}           69376& \cellcolor{pink}         4281984& \cellcolor{pink}       119221248& \cellcolor{pink}      2152806016\\
  3& \cellcolor{yellow}             0& \cellcolor{yellow}               0& \cellcolor{yellow}               0& \cellcolor{green}              96& \cellcolor{green}          631788& \cellcolor{green}       119221248& \cellcolor{pink}      8128381416  \\
  4& \cellcolor{yellow}             0& \cellcolor{yellow}               0& \cellcolor{yellow}               0& \cellcolor{yellow}               0& \cellcolor{green}             144& \cellcolor{green}         4247296& \cellcolor{green}      2152806016  \\
   5& \cellcolor{yellow}            0& \cellcolor{yellow}               0& \cellcolor{yellow}               0& \cellcolor{yellow}               0& \cellcolor{yellow}               0& \cellcolor{green}              96& \cellcolor{green}        24219184 \\
 6& \cellcolor{yellow}              0& \cellcolor{yellow}               0& \cellcolor{yellow}               0& \cellcolor{yellow}               0& \cellcolor{yellow}               0& \cellcolor{yellow}               0& \cellcolor{green}             144
          \end{array}
          \end{align*}
          \begin{align*}
\rotatebox[origin=c]{90}{\underline{$q_1=1$}}
\begin{array}{c|ccccccc}
      \mathdiagbox[width=0.7cm,height=0.5cm,innerleftsep=0.1cm,innerrightsep=0cm]{q_2}{q_3}   & 0 & 1 & 2 & 3 & 4 & 5 & 6   \\ \hline
  0&         0&               0&               0&               0&               0&               0&               0 \\
   1&      { \cellcolor{yellow} -2} &    \cellcolor{green}           96&    \cellcolor{green}         5384&    \cellcolor{green}        69376&      \cellcolor{green}     631788&      \cellcolor{green}    4247296& \cellcolor{green}        24219184 \\
  2& \cellcolor{yellow}             -4& \cellcolor{yellow}             288& \cellcolor{yellow}           -7968& \cellcolor{green}          227200& \cellcolor{pink}        47491408& \cellcolor{pink}      1531634496& \cellcolor{pink}     29149503744\\
  3& \cellcolor{yellow}                 -6& \cellcolor{yellow}             480& \cellcolor{yellow}          -15936& \cellcolor{yellow}          473472& \cellcolor{green}        -9517380& \cellcolor{green}      1720814976& \cellcolor{pink}    324341564352 \\
  4& \cellcolor{yellow}              -8& \cellcolor{yellow}             672& \cellcolor{yellow}          -23904& \cellcolor{yellow}          789120& \cellcolor{yellow}       -20298336& \cellcolor{green}       571788736& \cellcolor{green}     20359924864  \\
   5& \cellcolor{yellow}               -10& \cellcolor{yellow}             864& \cellcolor{yellow}          -31872& \cellcolor{yellow}         1104768& \cellcolor{yellow}       -30447504& \cellcolor{yellow}       945902400& \cellcolor{green}    -17554938576 \\
 6& \cellcolor{yellow}                          -12& \cellcolor{yellow}            1056& \cellcolor{yellow}          -39840& \cellcolor{yellow}         1420416& \cellcolor{yellow}       -40596672& \cellcolor{yellow}      1324263360& \cellcolor{yellow}    -26368736640
           \end{array}
          \end{align*}
               \begin{align*}
\rotatebox[origin=c]{90}{\underline{$q_1=2$}}
\begin{array}{c|ccccccc}
      \mathdiagbox[width=0.7cm,height=0.5cm,innerleftsep=0.1cm,innerrightsep=0cm]{q_2}{q_3}   & 0 & 1 & 2 & 3 & 4 & 5 & 6   \\ \hline
0&     0&               0&               0&               0&               0&               0&               0 \\
   1&           0&               0&               0&               0&               0&               0&               0 \\
  2& \cellcolor{yellow}             0& \cellcolor{yellow}               0& \cellcolor{green}             144& \cellcolor{green}           69376& \cellcolor{green}         4281984& \cellcolor{green}       119221248& \cellcolor{green}      2152806016\\
  3& \cellcolor{yellow}               -6& \cellcolor{yellow}             480& \cellcolor{yellow}          -15936& \cellcolor{yellow}          473472& \cellcolor{green}        -9517380& \cellcolor{green}      1720814976& \cellcolor{green}    324341564352\\
  4& \cellcolor{yellow}              -32& \cellcolor{yellow}            3360& \cellcolor{yellow}         -150480& \cellcolor{yellow}         4375040& \cellcolor{yellow}       -82286208& \cellcolor{yellow}      5273528256& \cellcolor{green}    -61597979328 \\
   5& \cellcolor{yellow}           -110& \cellcolor{yellow}           12960& \cellcolor{yellow}         -677376& \cellcolor{yellow}        23110528& \cellcolor{yellow}      -563180076& \cellcolor{yellow}     16877260800& \cellcolor{yellow}   -205787147392 \\
 6& \cellcolor{yellow}              -288& \cellcolor{yellow}           36960& \cellcolor{yellow}        -2128560& \cellcolor{yellow}        81523584& \cellcolor{yellow}     -2360300928& \cellcolor{yellow}     63452512704         & \cellcolor{yellow}  -1056669477312   \end{array}
          \end{align*}
              \begin{align*}
\rotatebox[origin=c]{90}{\underline{$q_1=3$}}
\begin{array}{c|ccccccc}
      \mathdiagbox[width=0.7cm,height=0.5cm,innerleftsep=0.1cm,innerrightsep=0cm]{q_2}{q_3}   & 0 & 1 & 2 & 3 & 4 & 5 & 6   \\ \hline
0&     0&               0&               0&               0&               0&               0&               0 \\
   1&           0&               0&               0&               0&               0&               0&               0 \\
  2&              0&               0&               0&               0&               0&               0&               0 \\
  3& \cellcolor{yellow}               0& \cellcolor{yellow}               0& \cellcolor{yellow}               0& \cellcolor{green}              96& \cellcolor{green}          631788& \cellcolor{green}       119221248& \cellcolor{green}      8128381416\\
  4& \cellcolor{yellow}             -8& \cellcolor{yellow}             672& \cellcolor{yellow}          -23904& \cellcolor{yellow}          789120& \cellcolor{yellow}       -20298336& \cellcolor{green}       571788736& \cellcolor{green}     20359924864 \\
   5& \cellcolor{yellow}            -110& \cellcolor{yellow}           12960& \cellcolor{yellow}         -677376& \cellcolor{yellow}        23110528& \cellcolor{yellow}      -563180076& \cellcolor{yellow}     16877260800& \cellcolor{yellow}   -205787147392 \\
 6& \cellcolor{yellow}              -756& \cellcolor{yellow}          105600& \cellcolor{yellow}        -6674400& \cellcolor{yellow}       267266592& \cellcolor{yellow}     -7543925760& \cellcolor{yellow}    201410856192& \cellcolor{yellow}  -4254252869280   \end{array}
          \end{align*}
                        \begin{align*}
\rotatebox[origin=c]{90}{\underline{$q_1=4$}}
\begin{array}{c|ccccccc}
      \mathdiagbox[width=0.7cm,height=0.5cm,innerleftsep=0.1cm,innerrightsep=0cm]{q_2}{q_3}   & 0 & 1 & 2 & 3 & 4 & 5 & 6   \\ \hline
0&     0&               0&               0&               0&               0&               0&               0 \\
   1&           0&               0&               0&               0&               0&               0&               0 \\
  2&              0&               0&               0&               0&               0&               0&               0 \\
  3&                0&               0&               0&               0&               0&               0&               0 \\
  4& \cellcolor{yellow}             0& \cellcolor{yellow}               0& \cellcolor{yellow}               0& \cellcolor{yellow}               0& \cellcolor{green}             144& \cellcolor{green}         4247296& \cellcolor{green}      2152806016 \\
   5& \cellcolor{yellow}            -10& \cellcolor{yellow}             864& \cellcolor{yellow}          -31872& \cellcolor{yellow}         1104768& \cellcolor{yellow}       -30447504& \cellcolor{yellow}       945902400& \cellcolor{green}    -17554938576 \\
 6& \cellcolor{yellow}             -288& \cellcolor{yellow}           36960& \cellcolor{yellow}        -2128560& \cellcolor{yellow}        81523584& \cellcolor{yellow}     -2360300928& \cellcolor{yellow}     63452512704& \cellcolor{yellow}  -1056669477312  \end{array}
          \end{align*}
\caption{Genus zero GV invariants $n_{q_1, q_2, q_3}^0$ for the geometry of Appendix \ref{app:branch_cuts}.
The cone of dual coordinates $\mathcal{T}$ is shown in pink, the region $\text{Vis}(\mathcal{T}_{\text{hyp}}) \setminus \mathcal{T}$ is shown in green, and the region $\mathcal{M}_\infty \setminus \text{Vis}(\mathcal{T}_{\text{hyp}}) $ is shown in yellow. There are nonzero GV invariants everywhere in the region $\text{Vis}(\mathcal{T}_{\text{hyp}})$ where BPS black holes are known to exist, in agreement with the lattice WGC. }
\label{GVextab}
          \end{table}

The only nop curve class is the one of charge $q_a = (1, 0 ,0)$, which shrinks to zero size at the boundary of the K\"ahler cone associated with the genus one Weyl reflection. By \eqref{w1}, we have
\begin{equation}
	w_1(\mathcal{K}) = \left\{ t^2 \geq - t^1 \geq 0 \,,~~t^3 \geq 0 \right\} \,.
\end{equation}
Correspondingly, the hyperextended K\"ahler cone is given by
\begin{equation}
	\mathcal{K}_{\text{hyp}} = \bigcup_{w\in \mathcal{W}_{\text{stable}}}  w(\mathcal{K})   =  \mathcal{K} \cup w_1(\mathcal{K}) = \left\{ t^2, t^3 \geq 0\,,~~ t^1 \geq - t^2   \right\}\,.
\end{equation}
In \S\ref{sec:modulispace}, we argued that the infinity cone $\mathcal{M}_\infty$ is dual to the hyperextended K\"ahler cone,
\begin{equation}
	\mathcal{M}_\infty  \overset{\eqref{eq:kinfty}}{=}  \mathcal{K}_{\text{hyp}}^\vee =
  \left \{ T_2 \geq T_1 \geq 0  \,,~ T_3 \geq 0 \right\} \,.
\end{equation}
We may compare this with the GV invariants of Table \ref{GVextab}. We have $n_{q_1, q_2, q_3}^0 \neq 0$ everywhere in the interior of $\mathcal{M}_\infty$, i.e. for all $q_2 > q_1 > 0$, $q_3 > 0$, implying that indeed $\mathcal{M}_\infty =  \mathcal{K}_{\text{hyp}}^\vee$.

The hyperextended K\"ahler cone maps to a region in dual coordinates of the form
\begin{equation}
	\mathcal{T}_{\text{hyp}} = \left \{ T_2 \geq T_1 \geq 0\,,~~ T_3 \geq 2 T_2 \right \}\,.
\end{equation}
Note that $\mathcal{T}_{\text{hyp}} \subset \mathcal{M}_{\infty}$, as required by the lattice WGC, since $\mathcal{T}_{\text{hyp}} \subseteq \mathscr{C}_{\text{BH}}$.

In fact, $\Vis(\mathcal{T}_{\text{hyp}}) \subseteq \mathscr{C}_{\text{BH}}$ as discussed in~\S\ref{sec:CBHcons}.  Upon examining the fully Weyl-extended moduli space, we encounter a surprise: as shown in Figure~\ref{fig:vis}, this moduli space is a \emph{branched cover} of the Weyl orbit $\mathcal{W}(\mathcal{T}_{\text{hyp}})$. In fact, passing repeatedly around the branch point, much of this moduli space turns out to lie outside $\mathcal{M}_\infty$. This sounds worrying from the perspective of the lattice WGC, but only $\Vis(\mathcal{T}_{\text{hyp}}) \subseteq \mathscr{C}_{\text{BH}}$ really matters for our analysis. Explicitly, this region is
\begin{align}
	\text{Vis}(\mathcal{T}_{\text{hyp}}) &= \mathcal{T}  ~\cup ~w_1(\mathcal{T})~\cup ~w_2(\mathcal{T})~ \cup~ w_2 \circ w_1 (\mathcal{T})~ \cup~ w_1 \circ w_2 (\mathcal{T})  )~ \cup~ w_1 \circ w_2 \circ w_1 (\mathcal{T}) \nonumber \\
	&= \left \{  T_2 \geq T_1 \geq 0\,,~ T_3 \geq \min( T_1 + T_2, 2 T_2 - T_1 ) \right\}   \,,
\end{align}
i.e., the (non-convex) vee-shaped region outlined by the green lines in Figure~\ref{fig:vis}. One can then verify that every GV invariant within this cone is non-zero up to the degree calculated --- see Table~\ref{GVextab} --- so the predictions of the lattice WGC are actually satisfied. The fact that the fully Weyl-extended moduli space eventually passes outside $\mathcal{M}_\infty$ is rather an indication that the BPS black hole solutions undergo wall crossing at the unstable Weyl flop. This would be very interesting to verify explicitly.

\newpage

\bibliography{refs}
\bibliographystyle{utphys}
\end{document}